\documentclass[aps,prx,twocolumn,superscriptaddress,nofootinbib,floatfix]{revtex4-2}

\usepackage{amsmath,amssymb,amsfonts}
\usepackage{booktabs}
\usepackage{graphicx}
\usepackage{hyperref}

\newcommand{\FF}{\ensuremath{\mathbb{F}}}
\newcommand{\code}[1]{\ensuremath{\left[\!\left[#1\right]\!\right]}}
\newcommand{\record}[2]{\ensuremath{\left[\!\left[#1\right]\!\right]^{\mathrm{#2}}}}
\newcommand{\FOM}{\ensuremath{\mathrm{FOM}}}
\newcommand{\supp}{\ensuremath{\operatorname{supp}}}

\begin{document}

\title{Large Language Model-Guided Discovery of Weight-Five Bivariate
Bicycle Codes}

\author{Juan Cruz-Benito}
\affiliation{IBM Research}

\date{5 October 2026}

\begin{abstract}
Building on our earlier program-evolution workflow guided by large language
models (LLMs), we study weight-five bivariate bicycle (BB) and perturbed
bivariate bicycle (PBB) codes.  The resulting catalogue contains 1,142 distinct
code proposals, including 1,081 nonbaseline proposals attributable to
LLM-generated programs.  Across the catalogue, we certify connected
Calderbank--Shor--Steane (CSS) realizations
$\code{96,4,10}$, $\code{140,6,10}$, and $\code{180,4,14}$.
A post-search comparison certifies seven imported
Lin--Pryadko archive constructions.  For leading parameter triples also
represented in that archive, we provide exact distance evidence, explicit
bivariate presentations, and verified component reductions.  A
basis-independent connectivity analysis identifies 409 of the 1,142 catalogue
entries as disconnected and shows that 73.1\% of the classes with exact
distance certificates contain repeated connected components.
Algebraic analysis organizes the connected CSS classes into order-3,
order-7, and order-15 cyclotomic-kernel strata.  The strongest exact
connected PBB parameter point is $\code{216,4,10}$, attained
by two distinct component classes.  Among the 936 distinct CSS proposals from
the LLM-guided campaign with a recorded positive distance, 816 (87.18\%) are
certified at $d\geq5$.  For comparison, three random-search controls each
sample 6,444 CSS code proposals uniformly without replacement, using the same
per-lattice and encoded-dimension sample counts as the LLM-guided campaign.
In these controls, 4,672--4,785 proposals (72.50--74.26\%) meet the same
criterion.  The LLM-guided campaign has the higher certified yield, while the
random controls cover more connected classes.  Together, these results extend
LLM-guided discovery to a more
constrained code family and provide a reproducible structural and
exact-distance account of its strongest candidates.
\end{abstract}

\maketitle

\section{Introduction}

Bivariate bicycle (BB) codes are a prominent family of
Calderbank--Shor--Steane (CSS) quantum low-density parity-check (qLDPC) codes.
The most extensively studied BB codes use two trinomials, yielding weight-six
stabilizer generators~\cite{bravyi2024high}.  Our previous work introduced and
benchmarked a program-evolution workflow guided by large language models (LLMs)
for BB and perturbed bivariate bicycle (PBB)
codes~\cite{cruzbenito2026evolutionary},
including comparisons with uniform random sampling and genetic algorithms,
and applied it to weight-six and higher-weight CSS and PBB codes.  Here we
extend that workflow to weight-five codes, a more constrained polynomial
family, and combine candidate discovery with systematic distance
certification, structural reduction, and algebraic classification.
``Weight five'' denotes the Pauli weight of each stabilizer generator.
For our regular binomial--trinomial ($2+3$) CSS constructions, the data-qubit
degree also equals five; circuit connectivity and gauge-check weight remain
distinct quantities.  We apply a stochastic LLM-guided program-evolution
search to bivariate codes with a
$2+3$ support split and block lengths $72\leq n\leq432$.  The CSS campaigns
optimize $kd^2/n$ across complementary profiles of small and large periodic
lattices, whereas
the PBB campaigns prioritize non-CSS candidates with $d\geq5$.  The targeted
search provides a finite-length view of these families.  The six CSS
distance-evaluation lattices contain
$13{,}656{,}867$ normalized binomial--trinomial pairs.  The logged
positive-distance output contains 936 distinct CSS proposals, of which 912
remain after the $d\leq2$ filter.
A three-replicate stratified-uniform control complements the general
uniform-random and genetic-algorithm comparisons in the earlier work by
probing transfer within the weight-five family.  A post-search comparison
establishes that our $\code{96,4,10}$ realization is equivalent to a
Lin--Pryadko archive construction over the cyclic group $C_{48}$ of order 48;
several additional results match archive parameters without an equivalence
claim.  For these overlaps, we provide exact certificates, explicit bivariate
realizations, and verified parent--component relations.
The resulting catalogue links the discovered proposals to distance evidence,
connected components, equivalence classes, algebraic structure, and
finite-length comparison frontiers.

\section{Related work and comparison scope}

Recent artificial-intelligence-assisted quantum error correction (QEC) code
discovery spans reinforcement-learning searches for
low-weight and decoder-tailored BB codes~\cite{he2025lowweight,sella2026decoder},
Bayesian optimization of BB codes~\cite{chengyu2026bayesian}, and LLM- or
agent-guided exploration of lifted-product and broader balanced-product
constructions and code--circuit--decoder
co-design~\cite{liu2026structured,yan2026omniqec,qian2026multiagent}.  The
present work belongs to this emerging methodological line while focusing on a
comparatively less explored structural regime: weight-five BB and related
qLDPC constructions.

In this regime, general weight-reduction methods can transform any qLDPC code
into one with stabilizer weight at most five, but need not preserve the bicycle
algebraic form~\cite{hastings2021weight}.  Because the finite-length overhead
depends on the construction, we include a transformed code as a weight-five
comparator only when its checks have been constructed explicitly and its
parameters certified.  Voss \emph{et al.} reported the trivariate
bicycle code $\code{30,4,5}$ and the additional weight-five examples
$\code{72,4,8}$ and $\code{96,4,8}$~\cite{voss2025multivariate}.  We include
them in the bivariate cohort because the third shift in their encoding
satisfies $z=xy$, reducing each presentation to an ordinary BB code---or,
equivalently, an abelian two-block group-algebra (2BGA) code.  Khesin and Lu
tabulated the weight-five mirror code
$\code{36,4,6}$, which is locally
Clifford equivalent to a CSS code~\cite{khesin2026mirror}.  Through the
mirror--2BGA correspondence, we likewise include its bivariate $2+3$
representation.  The $\code{30,4,5}$ construction tabulated by Khesin and Lu
is equivalent to the Voss code and is counted only once.

The generalized-bicycle and two-block group-algebra literature provides a
broader family-level comparison.  Panteleev and
Kalachev introduced generalized-bicycle (GB) codes over cyclic groups and
demonstrated their finite-length performance~\cite{panteleev2021degenerate}.
Lin and Pryadko exhaustively enumerated connected 2BGA presentations of stabilizer
weight at most eight, considering codes over abelian ambient groups through
block length $n=100$ and codes over nonabelian ambient groups through
$n=200$~\cite{lin2024twoblock}.  In their archive tuple $(G,k,d)$, $G$ is the
ambient group, $k$ the encoded dimension, and $d$ the distance; the associated
2BGA code has block length $n=2|G|$.  Their
accompanying machine-readable archive~\cite{linpryadko2bgaarchive} retains the
first construction encountered for each such tuple.  It is more detailed than
the parameter tables in the article and includes the weight-five constructions
used in our comparison below.  They estimated distances with the probabilistic
\texttt{DistRandCSS} routine from QDistRnd~\cite{pryadko2022qdistrnd}, using
$10^5$ trials.  We treat these archived values as randomized estimates and
distinguish them from exact distances in the comparison.

The archive serves as a post hoc comparison set and was not used during the
campaign.  Its exhaustive abelian range ends at block length $n=100$, below
the six searched abelian lattices, whose parent block lengths satisfy
$n\geq120$.  Moreover, retaining
one construction per $(G,k,d)$ tuple supports parameter-level comparison but
not same-lattice recall or code equivalence.  Sections~\ref{sec:css-results}
and~\ref{sec:pareto-comparison} report the equivalence and distance evidence
used for individual constructions.

The literature at adjacent check weights is much broader.  At weight six, it
includes the BB codes of Bravyi \emph{et al.}~\cite{bravyi2024high}, the exact
mirror code $\code{60,4,10}$~\cite{khesin2026mirror}, the coprime constructions
of Wang and Mueller~\cite{wang2024coprime}, and the generalized-toric
enumeration of Liang \emph{et al.}~\cite{liang2025generalized}.  Qian and Li
report coset-orbit balanced-product codes with distances certified by
mixed-integer linear programming (MILP), including
$\code{336,12,20}$ at overall weight six and $\code{224,22,16}$,
$\code{288,24,18}$, and $\code{378,32,19}$ at overall weight
eight~\cite{qian2026multiagent}.  From their published construction supports,
we assign these four to the same maximum-stabilizer-weight classes used here;
their overall-weight convention additionally bounds the total number of checks
incident on each qubit.  Other weight-eight comparators include the
covering-graph construction $\code{144,14,14}$~\cite{symons2025sequences} and
the catalogue from our previous search~\cite{cruzbenito2026evolutionary}.

For a broader benchmark, we use the August 28, 2026 repository snapshot
(accessed August 29, 2026) of the Unitary Foundation QEC
Challenge~\cite{unitaryfoundation2026qecchallenge}, a public
leaderboard of verifier-checked submissions.  We use its submitted family
labels as provenance metadata rather than equivalence classifications.  We
assign each entry a check weight equal to the
maximum nonzero row weight of its submitted $H_X$ and $H_Z$ matrices.  The
comparison set comprises all 97 entries labelled BB in that revision and the
five weight-five entries outside that family.  We analyze their evidence and
connectivity in Sec.~\ref{sec:pareto-comparison}.

Weight-five abelian 2BGA codes admit a three-dimensional local representation;
Ref.~\cite{arnault2026variant} derives the conditional finite-size upper bound
$d=O(n^{2/3})$.  Accordingly, the conventional metric $kd^2/n$ summarizes
finite-length performance.  For the literature comparison,
we construct separate exact Pareto frontiers by check weight for two pairs of
objectives: rate and distance, both maximized; and block length and $kd^2/n$,
with $n$ minimized and $kd^2/n$ maximized.  The plotted frontiers use exact
distances, with nonexact evidence retained in
the accompanying dataset.

\section{Search space and methods}
\label{sec:campaign}

\subsection{Code families and search objective}

For a periodic lattice with side lengths $(\ell,m)$, let
\begin{equation}
R_{\ell,m}=\FF_2[x,y]/(x^\ell-1,y^m-1).
\end{equation}
This is the binary group algebra of the lattice translation group
$C_\ell\times C_m$, where $C_r$ denotes the cyclic group of order $r$.  We
represent each polynomial by its
$\ell m\times\ell m$ binary circulant matrix.  A CSS BB code has check matrices
\begin{equation}
H_X=\begin{pmatrix}A&B\end{pmatrix},\qquad
H_Z=\begin{pmatrix}B^\mathsf{T}&A^\mathsf{T}\end{pmatrix},
\label{eq:bb}
\end{equation}
with block length $n=2\ell m$.  Commutativity of $R_{\ell,m}$ gives
$H_XH_Z^\mathsf{T}=AB+BA=0$.  We impose
\begin{equation}
\{|\supp(A)|,|\supp(B)|\}=\{2,3\}.
\label{eq:weight5}
\end{equation}
The $2+3$ and $3+2$ orientations differ only by exchanging the two qubit
sectors.  We exclude the $1+4$ split because a monomial block is a unit in
$R_{\ell,m}$, making both CSS check matrices full row rank; the resulting BB
code therefore encodes no logical qubits.

The non-CSS PBB ansatz from Ref.~\cite{cruzbenito2026evolutionary} is
\begin{equation}
H=\left(\begin{array}{cc|cc}
A&B&C&D\\
0&0&B^\mathsf{T}&A^\mathsf{T}
\end{array}\right),
\label{eq:pbb}
\end{equation}
where the vertical bar separates the symplectic $X$ and $Z$ coordinates.  We
require
\begin{equation}
\begin{aligned}
\supp(C)&\subseteq\supp(A), &
\supp(D)&\subseteq\supp(B),\\
(C,D)&\neq(0,0).
\end{aligned}
\end{equation}
Symplectic commutativity further requires
$AC^\mathsf{T}+BD^\mathsf{T}$ to be symmetric over $\FF_2$.  Support
containment ensures that the mixed rows have Pauli weight
$|\supp(A)\cup\supp(C)|+|\supp(B)\cup\supp(D)|=5$; each generator in the
second block row also has weight five.

For the CSS family in Eq.~\eqref{eq:bb}, every data qubit also has Tanner
degree $|\supp(A)|+|\supp(B)|=5$.  Stabilizer weight and qubit degree happen
to coincide for these regular constructions; they need not coincide for the
literature comparators or for general weight-reduction procedures.

We compare codes using the finite-length figure of merit (FOM)
\begin{equation}
\FOM=\frac{k d^2}{n},
\label{eq:fom}
\end{equation}
following our earlier work~\cite{cruzbenito2026evolutionary}.
Equation~\eqref{eq:fom} serves as a finite-length comparison metric without assuming
two-dimensional locality.  Its normalization echoes the two-dimensional
local-code tradeoff $kd^2=O(n)$~\cite{bravyi2010tradeoffs}.
For an idealized baseline of $k$ independent rotated surface-code patches,
each using $d^2$ data qubits, Eq.~\eqref{eq:fom} is exactly the ratio of
surface-code data qubits to BB-code data qubits at the same nominal $(k,d)$.
This data-qubit ratio isolates one finite-length resource measure; circuit and
hardware costs are evaluated separately where corresponding data are
available.

\subsection{Search protocol}

Four 300-iteration OpenEvolve campaigns~\cite{openevolve} evolved Python
functions that generate polynomial tuples across several lattices.  The CSS
fitness function favored high-FOM weight-five $2+3$ BB candidates, and the PBB
fitness function prioritized non-CSS candidates with $d\geq5$.  Each family
used complementary small- and large-lattice profiles and a staged protocol
that reserved distance calculations for promising candidates while sampling
low-, medium-, and high-$k$ bands.  An ensemble of Claude Opus 5 and GPT-5.6
Sol generated program variants, with model labels retained as provenance.
The small profile used stage-1 lattices
$(6,6),(6,9),(9,8)$ and stage-2 lattices $(6,10),(9,9),(10,9)$; the large
profile used $(12,9),(12,12),(15,12)$ and
$(16,12),(15,14),(18,12)$, respectively.  CSS distance evaluation used the
three stage-2 lattices of each profile, six in total.  For candidates that
passed Stage 1, PBB distance evaluation covered all six lattices of the
corresponding profile.

Generated tuples were normalized, deduplicated within each batch, filtered by
exact $\FF_2$ rank and the family constraints, and stratified into low-,
medium-, and high-$k$ bands before distance evaluation.  We call a
positive-distance record an \emph{observation} and a distinct normalized
$(\ell,m,A,B,C,D)$ tuple a \emph{proposal}.  Repeated observations can therefore
map to the same proposal.  Proposals with a proved value or reported endpoint
$d\leq2$ are excluded from the final catalogue.  The Supplemental Material gives the
per-campaign accounting.

\subsection{Distance certification and catalogue reduction}

Initial CSS ranking used belief propagation with ordered-statistics decoding
(BP--OSD), followed for the strongest candidates by combination-sweep OSD and
an exact-search attempt.  PBB candidates first underwent exhaustive
low-weight search and then size-dependent symplectic MILP and BP--OSD.  These
stages rank candidates and produce distance estimates or solver bounds;
exact-distance status is assigned using the criteria below.

We call $d$ exact only when exhaustive enumeration or an analytic argument
provides a lower bound matching a logical witness, or when all required MILP
problems terminate with proven optima.  Otherwise, an explicit logical witness
or a nonoptimal MILP incumbent supplies an upper bound $d\leq d_{\rm ub}$; a
decoder scalar whose operator was not retained is reported only as an estimated
upper endpoint.  The MILP uses the
Landahl--Anderson--Rice binary formulation~\cite{landahl2011color}, implemented
with \texttt{scipy.optimize.milp}~\cite{virtanen2020scipy} and HiGHS, with one
problem for each of the $2k$ logical objectives.  Every reported witness is
checked for weight, zero stabilizer syndrome, and nontrivial logical action.
Whenever weight-four exclusion and a weight-five witness jointly establish
$d=5$, we reconstruct explicit weight-five logical operators in both sectors.

Fixed follow-up rules selected retained CSS proposals with estimated
$\FOM\geq5$, all retained PBB-small proposals with $d=6$, and PBB-large
proposals evaluated by both MILP and BP--OSD.  The high-$k$ CSS tail received
componentwise analysis.  These rules concentrate certification on the strongest
decoder-ranked candidates.

To identify repeated presentations, we map every retained proposal to a
colored Tanner graph and canonicalize it with BLISS
~\cite{junttila2007engineering}.  CSS graph colors distinguish qubits,
$X$ checks, and $Z$ checks.  For PBB codes, separate vertices represent the
$X$ and $Z$ supports of each stabilizer row and a tying edge preserves their
pairing.  Equal canonical graph signatures define a presentation class under
qubit and stored-generator permutations, within which compatible distance
evidence is transferred.

For each CSS presentation class containing at least one proposal without a
certified direct lower bound, we
exhaustively searched both logical sectors for nontrivial operators of weight
at most four.  We enumerated all supports of at most two qubits and combined
pairs whose syndromes cancel, rejecting combinations that are stabilizers.  To
identify disconnected components independently of the chosen generators, we
row-reduced the binary symplectic
check matrix and grouped qubits that co-occur in its unique reduced row echelon
form.  This gives the finest decomposition of the stabilizer group into disjoint
qubit sets.  We then canonicalized one connected component of each parent and
used these component classes, rather than replicated parents, in the Pareto
comparison.

\subsection{Stratified-uniform transfer control}
\label{sec:uniform-control}

Our preceding paper benchmarks the LLM-guided workflow against uniform random
sampling and genetic algorithms in Sec.~VI F and Supplemental Appendix~B of
Ref.~\cite{cruzbenito2026evolutionary}.  Here we ask the narrower question of
how the workflow transfers to weight-five CSS codes.  On each of the six CSS
distance-evaluation lattices, three control replicates sample normalized pairs
$A=\{1,a\}$ and $B=\{1,b,c\}$ uniformly without replacement.  Sampling is
stratified by lattice over the $k=4$ and $k\geq6$ bands and continues until
each stratum matches the corresponding count of logged positive-distance
campaign observations.  The control therefore inherits the campaign's
lattice--band allocation and tests selection within strata, not allocation
among them.  Each selected pair follows the same BP--OSD cascade, followed by
an exhaustive search for nontrivial logical operators of weight at most four.
Exact evidence is transferred only between proposals in the same certified
presentation class.

The primary endpoints are the certified-$d\geq5$ fraction, connected
certified-class coverage, and highest-FOM campaign-certified class sampled.
The Supplemental Material summarizes campaign and solver budgets.  Full
prompts, configurations, canonicalization details, and validation artifacts
are available in the code repository~\cite{cruzbenito2026qcode}.

\section{Results}

\subsection{Catalogue certification and structure}
\label{sec:catalogue-results}

The four campaigns produced 13,923 positive-distance observations representing
1,297 distinct proposals.  After excluding 155 proposals with an exact value
or reported endpoint $d\leq2$, the retained catalogue contains 12,036 observations across
1,142 proposals: 912 CSS and 230 PBB.  Of these, 1,081 are nonbaseline
proposals attributable to LLM-generated programs.  The remaining 61 comprise
35 seed/null-only proposals, six without a retained attribution event, and 20
independently supplied fixed baselines; the baselines are excluded from the
1,081 even when evaluated by model-generated programs.  The Supplemental
Material gives the per-campaign accounting.

Distance certification began with the targeted follow-up summarized in
Table~\ref{tab:certification-outcomes}, which established exact distances for
106 of 175 retained targets.  Its 50 exact CSS results, together with six
retained high-$k$ component certificates, yielded exact distances for 56 CSS
proposals before the catalogue-wide low-weight search.

\begin{table}[tbp]
\centering
\caption{Certification outcomes for the targeted follow-up after
excluding proposals with $d\leq2$.  ``Nonexact'' combines a certified lower
bound with either a rigorous upper bound or a decoder estimate.}
\label{tab:certification-outcomes}
\begin{tabular}{lrrr}
\toprule
Campaign & Targets & Exact & Nonexact\\
\midrule
CSS-small & 16 & 12 & 4\\
CSS-large & 78 & 38 & 40\\
PBB-small & 30 & 30 & 0\\
PBB-large & 51 & 26 & 25\\
\midrule
Total & 175 & 106 & 69\\
\bottomrule
\end{tabular}
\end{table}

For the 96 prespecified CSS follow-up targets, including two later excluded at
$d=2$, the ratio of the initial estimated endpoint to the final reported
endpoint had medians 1.80 in the small profile and 2.08 in the large
profile, with maximum 14.5.  Class-level evidence also
materially tightens individual presentations: five presentations of one exact
$\code{432,4,10}$ class comprised one exact result at 10 and four decoder
estimates at 10, 20, 22, and 23; the exact result applies to all five.

The catalogue-wide low-weight search examined 492 CSS classes containing
proposals without direct lower-bound evidence.  It found two $d=2$ proposals,
which were excluded from the catalogue;
among retained entries it resolved 62 classes and 92 proposals at $d=3$ or 4,
and certified $d\geq5$ in 428 classes and 764 proposals.  Of the latter, 45
proposals also have explicit weight-five logical operators in both sectors and
are therefore exact.  Thus every retained CSS entry has a certified positive
lower bound.

The 1,142 retained proposals form 630 presentation classes: 535 CSS and 95
PBB.  Before the final presentation-class merge, 398 proposals have exact
row-level status.  Merging class evidence adds 38, giving 436 proposals in 227
exact classes; the remaining 403
classes retain certified lower bounds paired with either rigorous upper bounds
or decoder estimates.  No class contains conflicting distances or incompatible
bounds.

The basis-independent connectivity partition agrees with the stored-generator
Tanner partition for all 1,142 proposals.  It identifies 307 of 912 CSS
proposals (33.7\%) and 102 of 230 PBB proposals (44.3\%) as disconnected, for
409 of 1,142
overall (35.8\%).  At the class level, the corresponding counts are 206 of 535 CSS
(38.5\%) and 62 of 95 PBB (65.3\%), or 268 of 630 overall (42.5\%).  Among
those exact classes, 166 are disconnected (73.1\%): 108 of 146 CSS and 58 of 81
PBB.  For each disconnected proposal, the components have
equal size, their dimensions sum to the parent $k$, and their colored Tanner
graphs are pairwise isomorphic under BLISS canonical labelling.  Accordingly,
$m\times\code{n_0,k_0,d}$ reports the decomposition inferred from the parent
construction, not an independently generated smaller lattice presentation.
Explicit connected realizations for three CSS and four PBB component classes
are given in Secs.~\ref{sec:css-results} and~\ref{sec:pbb-results}.  Pareto
selection uses one representative of each component class, so replication
cannot create a frontier point.

\begingroup
\clubpenalty=10000
Componentwise analysis of seven high-$k$ CSS parents finds that each is a
direct sum of pairwise-isomorphic components.  One parent,
$18\times\code{24,4,2}$, is excluded by the $d\leq2$ rule; the other six are
retained exact catalogue entries tabulated in the Supplemental Material.  Full
component certificates are available in the repository
~\cite{cruzbenito2026qcode}.  In particular, all
four retained CSS proposals with $k\geq40$ are disconnected direct sums with
exact distances $d\leq5$.  More generally, 408 of the 409 disconnected
proposals have component dimension $k_0\in\{2,4,6,8\}$; the sole exception is
$\code{420,20,5}=2\times\code{210,10,5}$.  No retained connected CSS
proposal has $k\geq20$: the largest connected dimension is 16, attained
by the exact $\code{384,16,4}$ code, while the largest connected PBB dimension
is four.  Thus most of the catalogue's large parent dimensions arise through
replication.
\par
\endgroup

Repeating the canonical labelling after replacing each parent by one
connected component reduces the 630 parent classes to 608 component classes:
528 CSS and 80 PBB, with 206 exact and 402 nonexact classes.  This component
quotient is the catalogue used for comparisons between connected
constructions.  The shortest components with certified exact distances are a
PBB $\code{12,2,3}$ code and a CSS $\code{24,4,3}$ code, both with FOM 1.50.

The following subsections examine individual constructions.
Table~\ref{tab:headline-provenance} summarizes the provenance of the CSS, PBB,
and archive results discussed below.

\begin{table*}[t]
\centering
\caption{Provenance of selected exact results.  ``Parameter match'' records a
shared $(n,k,d)$; equivalence is listed where an explicit certificate is
available.}
\label{tab:headline-provenance}
\small
\begin{tabular}{@{}p{0.15\textwidth}p{0.41\textwidth}p{0.36\textwidth}@{}}
\toprule
Parameters & Role in this study & Prior-art relationship\\
\midrule
$\code{96,4,10}$ & Connected realization of a generated CSS parent & Equivalent to archive construction over $C_{48}$\\
$\code{140,6,10}$ & Connected realization of a generated CSS parent & Archive parameter match\\
$\code{180,4,14}$ & Generated directly as a connected CSS presentation & Archive parameter match\\
$\code{132,4,12}$ & Post-search archive construction & Distance certified here\\
$\code{216,4,10}$ & Two connected PBB component classes & ---\\
\bottomrule
\end{tabular}
\end{table*}

\subsection{CSS codes and leading candidates}
\label{sec:css-results}

Three exact disconnected parent codes have components that admit explicit
shorter BB presentations.  The
$\code{384,16,10}$ parent decomposes into four isomorphic components; the
connected BB presentation on $(\ell,m)=(16,3)$,
\[
 A=1+y+x^2y^2,\qquad B=1+x^3
\]
is BLISS-isomorphic to one component and therefore defines an exact
$\code{96,4,10}$ code.  Likewise, the three components of
$\code{420,18,10}$ admit the connected BB presentation on
$(\ell,m)=(5,14)$,
\[
 A=1+y^{13}+x^4y^{11},\qquad B=1+xy^7,
\]
giving $\code{140,6,10}$.  Their FOMs are 4.17 and 4.29.  In particular,
under the element ordering of the GAP (Groups, Algorithms, Programming)
computer-algebra system used by the Lin--Pryadko archive, the archived
$\code{96,4,10}$ construction over $C_{48}$ uses the group-algebra generator
pair $(1+z^3,1+z^{16}+z^2)$.  This is the pair in
Table~\ref{tab:algebraic}, up to exchanging the two polynomial blocks.  The
exact component certificate obtained here also certifies that archived
construction and identifies it with the explicit connected BB presentation on
$(\ell,m)=(16,3)$ given above.  This adds an explicit bivariate realization and
parent--component relation for that archive construction.
Relative to the Voss $\code{96,4,8}$ benchmark, the
$\code{96,4,10}$ parameter point has larger distance at the same $(n,k)$.

The exact $\code{180,8,9}=2\times\code{90,4,9}$ parent similarly gives the
connected BB presentation on $(\ell,m)=(5,9)$,
\[
 A=1+x^2y^8+x^3y^7,\qquad B=1+x^2y^6.
\]
Its component FOM is 3.60.  We retain this explicit construction and compare
it with the compiled benchmarks below.

Among the connected CSS codes in the present catalogue, $\code{180,4,14}$ has
the largest exact FOM, 4.356; $\code{180,4,13}$ provides another exact result,
with FOM 3.76.  Among connected CSS classes with nonexact distance, the largest
supported FOM ceiling is
$\code{432,4,5{\leq}d{\leq}23}$, at 4.90.  Two non-isomorphic singleton
classes tie next at $\code{420,4,5{\leq}d{\leq}22}$, with ceiling 4.61.  The
exhaustive low-weight search certifies the lower endpoints.  All three presentations were
included among the original 96 CSS targets.  The two length-420 upper endpoints
are nonoptimal MILP incumbents; an independently verified logical operator
tightens the length-432 upper bound from the MILP incumbent 24 to 23.
These interval classes are distinct from the exact $\code{432,4,10}$ and
$\code{420,4,5}$ presentation classes with the same respective $(n,k)$
values.
Table~\ref{tab:css} gives their generators alongside selected exact
constructions.  The supplement contains the compact catalogue, and the full
machine-readable catalogue is available in the repository.  At $n=120$, the connected $\code{120,4,10}$ and disconnected
$\code{120,16,5}=4\times\code{30,4,5}$ codes both have FOM 3.33; the
comparison uses the connected code.

\begin{table*}[t]
\centering
\caption{Selected connected weight-five CSS BB presentations.  Exact distances are
certified; interval rows combine a certified lower bound with a supported upper
endpoint, so the corresponding displayed FOMs are upper bounds.  M denotes exact
MILP certification over all logical sectors of the displayed presentation;
M+class transfers such a certificate from an isomorphic presentation in the
same presentation class; M+component denotes a connected BB realization
shown by BLISS canonicalization to be isomorphic to a component of an
MILP-certified parent.  B denotes a decoder estimate whose operator was not
retained, I a nonoptimal MILP incumbent, W an independently reconstructed and
verified logical operator, and L4 exhaustive exclusion through weight four.  Each row specifies
a presentation through its lattice and generators; matching $(n,k)$ values do
not imply equivalence.}
\label{tab:css}
\scriptsize
\begin{tabular}{lcllcl}
\toprule
Parameters & $(\ell,m)$ & $A(x,y)$ & $B(x,y)$ & $\FOM$ & Evidence\\
\midrule
$\code{90,4,9}$   & $(5,9)$ & $1+x^2y^8+x^3y^7$ & $1+x^2y^6$ & 3.60 & M+component\\
$\code{96,4,10}$  & $(16,3)$ & $1+y+x^2y^2$ & $1+x^3$ & 4.17 & M+component\\
$\code{120,4,10}$ & $(6,10)$ & $1+xy+x^5y^5$ & $1+y$ & 3.33 & M\\
$\code{140,6,10}$ & $(5,14)$ & $1+y^{13}+x^4y^{11}$ & $1+xy^7$ & 4.29 & M+component\\
$\code{180,4,13}$ & $(10,9)$ & $1+x^2y^3$ & $1+x^5y+x^7y^8$ & 3.76 & M\\
$\code{180,4,14}$ & $(10,9)$ & $1+x^2y^2+x^5y$ & $1+x^3y^3$ & 4.356 & M\\
$\code{432,4,10}$ & $(18,12)$ & $1+x^4y^3$ & $1+xy^{11}+x^2y^{10}$ & 0.93 & M+class\\
\midrule
$\code{420,4,5{\leq}d{\leq}22}$ & $(15,14)$ & $1+x^3y$ & $1+x^{10}y^6+x^{11}y^2$ & $\leq4.61$ & L4+I+B\\
$\code{420,4,5{\leq}d{\leq}22}$ & $(15,14)$ & $1+x^4y^3+x^8y^{11}$ & $1+x^3y^3$ & $\leq4.61$ & L4+I+B\\
$\code{432,4,5{\leq}d{\leq}23}$ & $(18,12)$ & $1+x^4y^3$ & $1+x^5y^{11}+x^{10}y$ & $\leq4.90$ & L4+I+W+B\\
\bottomrule
\end{tabular}
\end{table*}

The codes on the large-profile CSS lattices have block lengths beyond the
exhaustive abelian regime of Ref.~\cite{lin2024twoblock}.  The exactly certified
connected codes in this profile have parameters $\code{384,16,4}$,
$\code{384,4,4}$,
$\code{420,10,5}$, $\code{420,6,5}$, $\code{420,4,5}$,
$\code{420,6,3}$, $\code{432,4,10}$, and $\code{432,4,4}$.  The best FOM
among them is $0.93$ for $\code{432,4,10}$.

\subsection{Algebraic family classification}
\label{sec:algebraic-classification}

The exact connected CSS constructions classified below share a common
quotient-character structure.  Writing the lattice translation group
additively, let
$G=\mathbb Z_\ell\times\mathbb Z_m$ and let
$\phi:G\to\mathbb Z_q$ be a surjective homomorphism.  For a primitive
$q$th root of unity $\omega$ in an extension field of $\FF_2$, evaluate a
monomial by
\begin{equation}
 x^i y^j\longmapsto \omega^{\phi(i,j)}.
 \label{eq:quotient-character}
\end{equation}
If the two terms of the binomial have the same image and the three terms of
the trinomial sum to zero, both BB polynomials vanish on this character.  Its
Frobenius orbit is then a common-kernel sector.  A multiplicity-one order-three
orbit has degree two and contributes four logical dimensions; a
multiplicity-one order-seven orbit has degree three and contributes six.  In a
non-semisimple group algebra, a repeated common factor contributes according
to its multiplicity, so character vanishing alone does not determine $k$.  For
every selected construction below, direct rank calculations---and exact gcds
for the constructions over cyclic ambient groups---verify completeness and
multiplicity one.

This criterion is constructive.  Translate one trinomial exponent to the
identity.  For $q=3$, its other two exponents must map to residues $1$ and $2$,
while the binomial exponent difference lies in $\ker\phi$.  Thus a code in the
order-three stratum is generated by choosing a quotient map and placing the three
trinomial terms in the three distinct cosets of its kernel.  For $q=7$, choose
$\omega$ with $1+\omega+\omega^3=0$; the unordered residue pairs are
$(1,3)$, $(2,6)$, and $(4,5)$, with unit multiples of $\phi$ covering the
conjugate primitive factor.  The same kernel condition applies to the
binomial.  These rules generate the order-three and order-seven strata before
any distance calculation; direct rank calculation still determines
multiplicity and the resulting $k$.

No analogous order-five stratum can exist.  Because $2$ has multiplicative
order four modulo five, a primitive fifth root $\omega$ has minimal polynomial
$\Phi_5(z)=1+z+z^2+z^3+z^4$ over $\FF_2$.  Under a surjective
$\phi:G\to\mathbb Z_5$, the trinomial evaluates to $t(\omega)$, where
$t(z)=z^{r_1}+z^{r_2}+z^{r_3}$ is reduced in $\FF_2[z]/(z^5-1)$ and
$r_i\in\mathbb Z_5$ are the images of its three exponents.  If two residues
coincide, those terms cancel and $t$ is a single monomial; otherwise $t$ has
exactly three terms.  In either case $t$ is nonzero of degree at most four, so
$t(\omega)=0$ would force $t=\Phi_5$, which has five terms.  Hence the trinomial
of a weight-five $2+3$ code never vanishes on an order-five character, for any
lattice.  The same argument excludes every prime $q\geq5$ for which $2$ is a
primitive root modulo $q$ ($q=5,11,13,\ldots$), since $\Phi_q$ is then
irreducible with $q$ terms.  The argument does not apply to the orders that
occur: $\Phi_3$ is itself a trinomial, and $\Phi_7$ and $\Phi_{15}$ each
factor over $\FF_2$ into two trinomials.

There is also a simple restriction specific to the $2+3$ support split.  For
an abelian BB code, $k/2$ is the dimension of the common kernel of the two
regular-representation blocks.  If that kernel were one-dimensional, its
translation action would be trivial because
$\operatorname{GL}(1,\FF_2)=\{1\}$, so it would be spanned by the all-ones
vector.  The even-weight binomial annihilates that vector, whereas the
odd-weight trinomial does not.  Hence a nontrivial code in this family cannot
have $k=2$.  Independently, $k=2\dim(\ker A\cap\ker B)$ is even and therefore
excludes $k=3$.  Its smallest possible nonzero dimension is consequently
$k=4$, explaining why the campaign's $k\in\{2,3\}$ sampling quota was empty.

When $\gcd(\ell,m)=1$, the ambient group is cyclic.  A change of generator
based on the Chinese remainder theorem (CRT) identifies the code with a
generalized-bicycle code over $\FF_2[z]/(z^{L}-1)$, where $L=\ell m$.  If its
two polynomials are $a(z)$ and $b(z)$, the GB/coprime-BB dimension
formula~\cite{panteleev2021degenerate,wang2024coprime} is
\begin{equation}
 k=2\deg h,\qquad h=\gcd(a,b,z^L-1).
 \label{eq:cyclic-kernel-dimension}
\end{equation}
Table~\ref{tab:algebraic} identifies five exact constructions over cyclic
ambient groups as members of the known cyclic GB/coprime-BB subclass.  The two
constructions over noncyclic ambient groups share the same order-three quotient
mechanism, connecting all seven through a common design rule.

For connected $2+3$ presentations over cyclic ambient groups, $h$ is
necessarily squarefree.  Translating each polynomial by a monomial changes
neither $h$ nor connectivity, so write the binomial as $1+z^e$ and the
trinomial as $1+z^f+z^g$ with $0<f<g<L$.  Suppose an irreducible $p$ satisfies
$p^2\mid h$.  Since $h=\gcd(z^L-1,\,1+z^e,\,1+z^f+z^g)$, $p^2$ divides each of
the three polynomials individually, hence so does $p$; a repeated
irreducible factor of a polynomial also divides its formal derivative, so $p$
divides the formal derivatives of $z^L-1$, $1+z^e$, and $1+z^f+z^g$, and
$p\neq z$ because $z\nmid z^L-1$.  The first two conditions force $L$ and $e$
to be even.  If exactly one of $f$ and
$g$ were odd, $p$ would divide a monomial; if both were odd, $p$ would divide
$z^{f-1}(1+z^{g-f})$, hence $1+z^{g-f}$, and therefore
$1+z^f(1+z^{g-f})\equiv1\pmod p$.  Both cases are impossible, so $e$, $f$,
$g$, and $L$ are all even.  Both translated supports then lie in the
index-two subgroup $2\mathbb Z_L$, each check acts on a single coset of
$2\mathbb Z_L$ in both blocks, and the Tanner graph splits into two parity
components, contradicting connectivity.

\begin{table*}[t]
\centering
\caption{Algebraic classification of the exact connected CSS constructions.
The cyclic forms use a CRT generator $z$; $C_r$ denotes the cyclic group of
order $r$.  The ordering of $a$ and $b$ matches
the corresponding presentation in Table~\ref{tab:css}, which need not be the
representative selected for the supplemental class summary.  For
noncyclic ambient groups, the last column gives an explicit quotient character.}
\label{tab:algebraic}
\scriptsize
\setlength{\tabcolsep}{4pt}
\begin{tabular}{lcll}
\toprule
Parameters & Ambient group & Cyclic form $(a(z),b(z))$ & Complete common-kernel sector\\
\midrule
$\code{90,4,9}$ & $C_{45}$ & $(1+z^{17}+z^{43},\ 1+z^{42})$ & $h=z^2+z+1$ (order 3)\\
$\code{96,4,10}$ & $C_{48}$ & $(1+z^2+z^{16},\ 1+z^3)$ & $h=z^2+z+1$ (order 3)\\
$\code{120,4,10}$ & $C_6\times C_{10}$ & --- & $\phi(i,j)=i\pmod 3$\\
$\code{140,6,10}$ & $C_{70}$ & $(1+z^{39}+z^{55},\ 1+z^{21})$ & $h=z^3+z^2+1$ (order 7)\\
$\code{180,4,13}$ & $C_{90}$ & $(1+z^{12},\ 1+z^{17}+z^{55})$ & $h=z^2+z+1$ (order 3)\\
$\code{180,4,14}$ & $C_{90}$ & $(1+z^2+z^{55},\ 1+z^3)$ & $h=z^2+z+1$ (order 3)\\
$\code{432,4,10}$ & $C_{18}\times C_{12}$ & --- & $\phi(i,j)=j\pmod 3$\\
\bottomrule
\end{tabular}
\end{table*}

The chosen lattice profile strongly influences which low-order quotient
characters can occur; the observed pattern is not a consequence of weight five
alone.  All six CSS
distance-evaluation translation groups, of orders 60, 81, 90, 192, 210, and
216, admit an
order-three quotient, whereas only the order-210 group admits an order-seven
quotient.  Orders 60, 90, and 210 admit an order-15 quotient; the observed
order-15 classes occur only at orders 90 and 210.

For the support-subgroup comparisons below, let
\begin{equation}
k_S=\dim\ker A+\dim\ker B-\dim\ker(AB).
\end{equation}
The substantive relation is
\begin{equation}
k_S\leq\dim(\ker A\cap\ker B)=k/2,
\label{eq:ks-bound}
\end{equation}
with equality if and only if $\ker(AB)=\ker A+\ker B$.  Equality holds in
the semisimple case but can fail when $|G|$ is even.

The two selected exact classes presented over noncyclic ambient groups exhibit
a second structural pattern.
After translating one term of each support to the identity, the supports of both
polynomials generate cyclic subgroups even though the ambient group is
noncyclic.  For $\code{120,4,10}$, the two directions may be chosen as
$u=(1,1)$ of order 30 and $v=(0,1)$ of order 10 in
$C_6\times C_{10}$, with $u^6=v^6$ and an intersection of order five.  For
$\code{432,4,10}$, $u=(4,3)$ and $v=(1,11)$ both have order 36 in
$C_{18}\times C_{12}$, with $u^6=v^6$ and an intersection of order six.
For $\code{120,4,10}$, the common subgroup
$N=\langle u\rangle\cap\langle v\rangle$ satisfies $N\cong C_5$; for
$\code{432,4,10}$, it satisfies $N\cong C_6$.  For $\code{120,4,10}$, both
cyclic support-group extensions
over $N$ split, and the quasi-abelian lifted-product reduction of Lin and
Pryadko~\cite{lin2024twoblock} identifies it with a hypergraph-product code
over the binary group algebra $\FF_2[C_5]$.  For $\code{432,4,10}$, the
corresponding $C_{36}$
extensions over $C_6$ do not split.  In the support-subgroup rank decomposition,
its $k_S$ lower bound vanishes even though the common-kernel dimension is two,
so Eq.~\eqref{eq:ks-bound} is strict.  The normal cyclic overlap alone therefore does not
justify assigning this code the same lifted-product reduction.

We refer to this broader pattern as the noncyclic two-direction
cyclic-support stratum.  It contains 28 of the 114 noncyclic connected CSS
presentation classes; the other 86 do not have cyclic support subgroups in any
retained presentation.  The split or nonsplit character further divides this
stratum: 18 of the 28 classes are split and 10 are nonsplit, with the same
classification across all retained presentations.  Eight of the 10 nonsplit
classes, including $\code{432,4,10}$, form a BB stratum
characterized by nonsplit cyclic overlap and a strict bound:
$k=4$ and $k_S=0<k/2$.  The remaining two saturate
Eq.~\eqref{eq:ks-bound}, so nonsplitting alone does not imply strictness.  The
$\code{120,4,10}$ and
$\code{432,4,10}$ codes are the two high-distance exact members highlighted
here.  Seven of the 28 classes are exact (five at $d=4$ and these two at
$d=10$), compared with nine of the other 86 noncyclic classes (two at $d=3$,
four at $d=4$, and three at $d=5$).  Certified lower bounds of at least five
occur in 23 of 28 and 80 of 86 classes, respectively.  Because certification
targeted selected candidates and nonexact distances are censored, we interpret
these counts as structural coverage statistics for the stratum rather than as
performance estimates.

Low-order quotient characters appear throughout the sampled lattices, not only
among the leading codes.  Enumerating every surjective homomorphism to
$\mathbb Z_3$, $\mathbb Z_5$, $\mathbb Z_7$, and $\mathbb Z_{15}$ yields an
order-three certificate in 298 of the 329 connected CSS
presentation classes and an order-seven certificate in 21; two classes have both.
As the argument above requires, no class has an order-five certificate.
Thus 317 of 329 classes lie in those
two
low-order strata.  The remaining 12 are precisely classes with cyclic
presentations and one of
the two primitive order-15 quartic factors.  Hence all 329 connected CSS
classes fall into order-three, order-seven, or order-15 cyclotomic-kernel
strata.  No order-15 class appears among the selected
high-distance exact constructions.  Of the 215 connected classes with a
coprime lattice presentation, the exact cyclic gcd, including multiplicity, is
$z^2+z+1$ for 182 classes, one of the two primitive order-seven cubics for 19,
one of the two primitive order-15 quartics for 12, and the product of the
order-three factor with an order-seven cubic for two.  The remaining 114
classes are presented over noncyclic ambient groups, and all have an order-three
quotient certificate.  Every computed cyclic gcd is squarefree, as the
connectivity argument above requires; because the gcds were computed with
multiplicity, this agreement is an independent check rather than an assumption
of Eq.~\eqref{eq:cyclic-kernel-dimension}.
The calculation can be reproduced using code in the qcode-discovery
repository~\cite{cruzbenito2026qcode}.

The same low-order quotient structure appears in earlier weight-five
constructions.  Substituting $z=xy$ in the three Voss constructions gives
ordinary bivariate BB codes, each with an order-three quotient character; the
CSS presentation of the $\code{36,4,6}$ Khesin--Lu mirror code has the same
property.  We therefore use ``low-order quotient character'' and ``cyclotomic
kernel'' as a structural taxonomy within the known abelian 2BGA/BB family.
The order-seven $\code{140,6,10}$ code realizes the factor-first coprime-BB
approach at weight five.  The leading interval classes---one
$\code{432,4,5\leq d\leq23}$ class and two non-isomorphic
$\code{420,4,5\leq d\leq22}$ classes---are further members of the order-three
stratum; exact resolution would enable a performance comparison.

\subsection{Pareto comparison across check weights}
\label{sec:pareto-comparison}

\begin{figure*}[t]
\centering
\includegraphics[width=\textwidth]{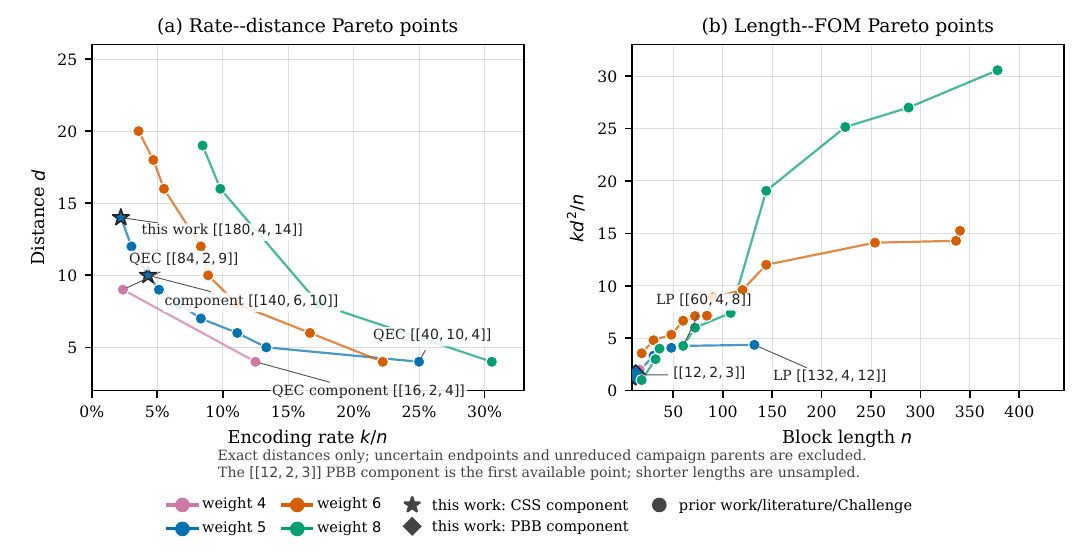}
\caption{Exact-distance Pareto frontiers within the compiled dataset, grouped
by maximum stabilizer weight after excluding $d\leq2$.  Panel (a) shows the
nondominated rate--distance frontier; panel (b) shows the idealized data-qubit
ratio $kd^2/n$ relative to $k$ independent rotated surface-code patches as
block length increases.  Stars and diamonds denote component-class points
originating from CSS and PBB codes in this study, respectively; circles denote
prior constructions and Challenge entries; color denotes maximum stabilizer
weight.
The cohorts are noncumulative, and the compiled dataset is not an exhaustive
census of the corresponding design spaces.
The $\code{12,2,3}$ label marks the first available point in the compiled
weight-five series; shorter lengths are currently unsampled.  Lines are guides
to the eye.  Every current or
prior campaign entry with stored polynomial supports is reduced to one
translation-connected component before frontier selection, and all Challenge
entries receive the basis-independent connectivity test.}
\label{fig:pareto-comparison}
\end{figure*}

Figure~\ref{fig:pareto-comparison} compares exact per-weight Pareto frontiers in
the compiled dataset: this study's weight-five catalogue, our previous
weight-six and weight-eight results, published benchmarks, and the Challenge
comparison set defined above.  Each included construction is assigned to a
cohort by its maximum stabilizer weight, producing disjoint fixed-weight cohorts rather than
cumulative ``weight at most $w$'' envelopes.  Their ordering at short block
lengths is interpreted in that fixed-weight sense.  Only exact distances enter
the figure.  For this
study, the figure uses one connected representative per component class
rather than unreduced parent presentations.  For earlier-campaign codes with
polynomial supports, a translation-subgroup test computes the index in
$\mathbb{Z}_{\ell}\times\mathbb{Z}_m$ of the subgroup generated by differences
within $\operatorname{supp}(A)\cup\operatorname{supp}(C)$ and within
$\operatorname{supp}(B)\cup\operatorname{supp}(D)$, with $C=D=0$ for CSS
codes; this index gives the number of translation components of the stored
presentation.  Every Challenge entry is rebuilt from its submitted $H_X,H_Z$ matrices
and reduced using the same invariant row-space decomposition as the present
catalogue.  Certified intervals, decoder estimates, witness-only ceilings, and
disconnected parents remain available in the accompanying comparison dataset.

Challenge entries without exact certificates are treated only as
witness-based upper bounds.  When different sources report identical
parameters, the comparison plots that parameter point once using the strongest
available evidence; this aggregation does not assert code equivalence.
The component reduction changes several entries used to construct the frontiers: a prior weight-six
$\code{144,16,8}$ parent reduces to $2\times\code{72,8,8}$, a prior
weight-eight $\code{144,44,4}$ parent to $2\times\code{72,22,4}$, and a prior
weight-eight PBB $\code{180,12,8}$ parent to
$3\times\code{60,4,8}$.  The corresponding components, rather than their
replicated parents, enter the frontiers.  The Challenge analysis similarly reduces
$\code{96,12,4}$ and $\code{192,24,4}$ to six and twelve copies,
respectively, of the same connected $\code{16,2,4}$ component.
Table~\ref{tab:pareto-points} lists every exact Pareto point in each panel so
that the figure need not be densely labelled.

\begin{table*}[t]
\centering
\caption{Exact-distance points defining the per-weight frontiers in the
compiled dataset shown in Fig.~\ref{fig:pareto-comparison}, listed in plot order.  Source superscripts
are O (this work, connected CSS catalogue proposal), O$_c$ (this work,
CSS component-class point, with a separately supplied explicit polynomial
realization when available), O$_p$ (this work, PBB component-class point), Q
(connected QEC Challenge entry), Q$_c$ (QEC Challenge component),
V (Voss \emph{et al.}~\cite{voss2025multivariate}),
K (Khesin--Lu~\cite{khesin2026mirror}),
B (Bravyi \emph{et al.}~\cite{bravyi2024high}),
LP (Lin--Pryadko archive construction, with distance independently certified
here),
L (Liang \emph{et al.}~\cite{liang2025generalized}), P (our previous
work~\cite{cruzbenito2026evolutionary}), QL
(Qian--Li~\cite{qian2026multiagent}), and
S (Symons \emph{et al.}~\cite{symons2025sequences}).}
\label{tab:pareto-points}
\scriptsize
\setlength{\tabcolsep}{3pt}
\begin{tabular}{ccp{0.82\textwidth}}
\toprule
Weight & Panel & Pareto points, in plot order\\
\midrule
4 & (a) & \record{16,2,4}{Q_c}, \record{84,2,9}{Q}\\
4 & (b) & \record{16,2,4}{Q_c}\\
5 & (a) & \record{180,4,14}{O}, \record{132,4,12}{LP},
  \record{140,6,10}{O_c}, \record{78,4,9}{LP},
  \record{60,4,8}{LP}, \record{48,4,7}{LP}, \record{36,4,6}{K},
  \record{30,4,5}{V}, \record{40,10,4}{Q}\\
5 & (b) & \record{12,2,3}{O_p}, \record{30,4,5}{V}, \record{36,4,6}{K},
  \record{48,4,7}{LP}, \record{60,4,8}{LP}, \record{132,4,12}{LP}\\
6 & (a) & \record{336,12,20}{QL}, \record{340,16,18}{L},
  \record{254,14,16}{L}, \record{144,12,12}{B}, \record{90,8,10}{B},
  \record{72,8,8}{P}, \record{72,12,6}{B}, \record{18,4,4}{Q}\\
6 & (b) & \record{18,4,4}{Q}, \record{30,4,6}{Q},
  \record{48,4,8}{Q}, \record{60,4,10}{K}, \record{72,8,8}{P},
  \record{84,6,10}{Q},
  \record{90,8,10}{B}, \record{120,8,12}{L}, \record{144,12,12}{B},
  \record{254,14,16}{L}, \record{336,12,20}{QL}, \record{340,16,18}{L}\\
8 & (a) & \record{378,32,19}{QL}, \record{224,22,16}{QL},
  \record{288,50,8}{P},
  \record{72,22,4}{P}\\
8 & (b) & \record{18,2,3}{P}, \record{32,6,4}{P},
  \record{36,4,6}{P}, \record{60,4,8}{P}, \record{72,12,6}{P},
  \record{108,8,10}{P}, \record{144,14,14}{S},
  \record{224,22,16}{QL}, \record{288,24,18}{QL},
  \record{378,32,19}{QL}\\
\bottomrule
\end{tabular}
\end{table*}

Exhaustive enumeration independently verifies the reported distances of the
Voss $\code{30,4,5}$ code and the Khesin--Lu $\code{36,4,6}$ mirror/2BGA
code; their FOMs are 3.33 and 4.00, respectively
~\cite{voss2025multivariate,khesin2026mirror}.  Voss also reports
$\code{72,4,8}$ and $\code{96,4,8}$, with FOMs 3.56 and 2.67; these remain
literature-reported values.  The connected constructions from this study
extend the range of distances with exact certificates in this comparison.
This is a check-weight-constrained comparison.  For unrestricted context,
CodeTables gives the exact optimum $\code{12,2,4}$ and known constructions
$\code{30,4,8}$, $\code{36,4,9}$, $\code{48,4,11}$, and
$\code{60,4,14}$~\cite{grassl2026codetables}.  These provide unrestricted
reference points for the distance gap associated jointly with qLDPC sparsity,
the bicycle algebraic form, and weight five at matched $(n,k)$.
In the post-search comparison, MILP over all logical sectors certified five
Lin--Pryadko archive constructions over cyclic ambient groups, with parameters
$\code{48,4,7}$, $\code{60,4,8}$, $\code{66,4,8}$,
$\code{78,4,9}$, and $\code{84,4,9}$.  The exact component certificate and
explicit CRT equivalence certify the archived $\code{96,4,10}$ construction
over $C_{48}$, while MILP certifies the archived $\code{132,4,12}$ construction
over a nonabelian ambient group.  These results preserve the original
attribution while allowing the constructions to enter the exact-distance
comparison.  The resulting exact length--FOM envelope reaches 4.27 at
$\code{60,4,8}$ and
4.364 at $\code{132,4,12}$.  The latter narrowly exceeds our
$\code{180,4,14}$ code's FOM of 4.356.  Thus,
$\code{132,4,12}$ has the highest FOM on the exact length--FOM envelope, while
$\code{140,6,10}$ and $\code{180,4,14}$ add points to the exact
rate--distance frontier.
The Voss $\code{72,4,8}$ value remains in the comparison dataset as a
literature-reported estimate; the exact-only frontier uses certified points.
At the reported value, the Voss point is dominated by the exact Lin--Pryadko
$\code{60,4,8}$ archive point, while the exact Lin--Pryadko $\code{78,4,9}$
archive point dominates our derived $\code{90,4,9}$ component.  Both dominated
points remain available in the comparison data.

The archive also contains constructions over nonabelian ambient groups with the
same parameter triples as our $\code{120,4,10}$, $\code{140,6,10}$,
$\code{180,4,13}$, and $\code{180,4,14}$ codes.  We establish explicit code
equivalence for $\code{96,4,10}$ above; for these four constructions, the
archive relationship is currently a parameter match.  Our $\code{432,4,10}$
construction lies beyond the archive's enumeration over nonabelian ambient
groups through block length $n=200$.

The archive also reports constructions over nonabelian ambient groups with
parameters $\code{168,4,14}$ (FOM 4.67), $\code{180,4,15}$ (5.00), and
$\code{192,4,16}$ (5.33).  Their archived distances were obtained using a
randomized routine.
Our all-logical MILP attempts found matching weight-14, weight-15, and weight-16
witnesses, respectively, and proved optimality for 1 of 8, 0 of 8, and 1 of 8
logical objectives within the per-logical time limit.  Every MILP run found an
incumbent at the reported weight, establishing the three values as upper bounds
in the comparison dataset.  If all three
reported distances were eventually certified, $\code{168,4,14}$ would dominate
the campaign's $\code{180,4,14}$ point, while $\code{140,6,10}$ would remain on
the rate--distance frontier.

Analysis of their support subgroups further separates these three
presentations.  The archived presentation with $(n,k)=(180,4)$ and upper bound
$d\leq15$ has a central overlap $N\cong C_{15}$ with split support-group extensions
and admits the quasi-abelian
reduction to a hypergraph-product code over $\FF_2[C_{15}]$; here
$k_S=k/2=2$.  For the archived presentations with $(n,k)=(168,4)$ and
$(192,4)$ and upper bounds 14 and 16, respectively, the central overlaps are
isomorphic to $C_{14}$ and $C_8$, respectively; one support-group extension is
nonsplit in each case, and direct rank calculations
give the strict inequality $k_S=0<k/2=2$.  This presentation-level
classification is structural; distance certification and equivalence to an
abelian BB code are separate questions.

The exact $\code{132,4,12}$ construction has ambient group $C_{11}\times S_3$,
where $S_3=\langle r,s\mid r^3=s^2=1,\ srs=r^{-1}\rangle$ is the symmetric
group on three letters.  Its binomial and trinomial supports generate cyclic
groups $C_{22}$ and $C_{33}$ with common central overlap $C_{11}$; modulo that
overlap, the corresponding group-algebra elements are $1+s$ and $1+r+r^2$ in
$\FF_2[S_3]$.  It is therefore a quasi-abelian
lifted-product code over $\FF_2[C_{11}]$, but no abelian 2BGA presentation has
been established.  Under the
idealized comparison of Eq.~\eqref{eq:fom}, the ratios of data qubits in $k$
independent rotated surface-code patches to those in the $\code{96,4,10}$,
$\code{140,6,10}$, and $\code{180,4,14}$ codes are 4.17, 4.29, and 4.356,
respectively.  These ratios compare data-qubit counts at the same nominal
$(k,d)$; a total-hardware comparison additionally requires circuit-level
resources.

Higher-weight checks can yield substantially larger exact FOMs.  Exact
weight-six reference points include the mirror code
$\code{60,4,10}$ with FOM 6.67~\cite{khesin2026mirror},
$\code{288,12,18}$ with FOM 13.5~\cite{bravyi2024high},
$\code{336,12,20}$ with FOM 14.29~\cite{qian2026multiagent}, and
$\code{340,16,18}$ with FOM 15.25~\cite{liang2025generalized}.  Within the
compiled dataset, the Qian--Li $\code{336,12,20}$ point replaces
$\code{294,10,20}$ on the exact weight-six rate--distance frontier and joins
the length--FOM frontier.  The weight-six
$\code{360,12,{\leq}24}$ benchmark is currently known only through the ceiling
$\FOM\leq19.20$~\cite{bravyi2024high}.  At
weight eight, the MILP-certified $\code{144,14,14}$ has exact FOM
19.06~\cite{symons2025sequences}.  Within the compiled dataset, Qian and Li
extend the exact length--FOM frontier through $\code{224,22,16}$,
$\code{288,24,18}$, and
$\code{378,32,19}$, with FOMs 25.14, 27.00, and 30.56,
respectively~\cite{qian2026multiagent}; the first and third also enter the
rate--distance frontier.  These four Qian--Li points enter at their published
parameters and exact MILP distances; their component structure was not
independently evaluated here.  Exact codes from our previous work include
$\code{144,12,12}$ with FOM 12.00 and $\code{288,50,8}$ with FOM
11.11~\cite{cruzbenito2026evolutionary}.
Repeated parameter triples in different maximum-weight cohorts correspond to
distinct presentations; parameter equality here does not imply equivalence.
This applies to
$\code{36,4,6}$ at weights five and eight, $\code{60,4,8}$ at weights five
and eight, $\code{72,12,6}$ at weights six and eight, and
$\code{144,12,12}$ at weights six and eight.  Two additional short weight-eight
Pareto points result from the same component reduction: the prior parents
$\code{144,16,3}$ and $\code{288,54,4}$ reduce to
$8\times\code{18,2,3}$ and $9\times\code{32,6,4}$, respectively.
Matched block lengths make the tradeoff associated with lower check weight
especially clear.  Liang \emph{et al.} exactly certify weight-six
$\code{96,4,12}$ and $\code{140,6,14}$ codes, compared with our weight-five
$\code{96,4,10}$ and $\code{140,6,10}$.  At $n=180$, their exact
$\code{180,8,16}$ improves both $k$ and $d$ over our $\code{180,4,14}$.
These length-matched comparisons clarify the stabilizer-weight tradeoff.
As a scheduling proxy, the combined CSS incidence graph of every $2+3$
construction here is bipartite and 5-regular, so K\H{o}nig's line-coloring
theorem fixes the graph-theoretic minimum at five conflict-free check--data
interaction layers.  The analogous minimum is six for regular $3+3$
weight-six BB codes.  These minima are not fault-tolerant circuit depths; for
example, Ref.~\cite{bravyi2024high} gives a seven-round extraction schedule
for weight-six BB codes.  Ancilla preparation and measurement, hook-error
ordering, routing, and hardware constraints can add further layers.
For the family-relevant scale suggested by the conditional
$d=O(n^{2/3})$ bound, the ratios $d/n^{2/3}$ for
$\code{96,4,10}$, $\code{140,6,10}$, and $\code{180,4,14}$ are 0.477,
0.371, and 0.439, respectively.  The corresponding ratios for the exact
weight-five comparators $\code{30,4,5}$, $\code{60,4,8}$, and
$\code{132,4,12}$ are 0.518, 0.522, and 0.463.  These are finite-size
descriptors and do not establish an asymptotic trend.

Of the 97 Challenge entries labelled bivariate bicycle, 55 have exact
certificates and 42 have witness-only upper bounds.  Under our row-weight
definition, five have weight four, 76 weight six, 10 weight eight, one weight
10, and five weight greater than 25; none has weight five.  The weight-four
entries are the exact $\code{84,2,9}$, $\code{96,12,4}$, and
$\code{192,24,4}$ codes and the witness-only $\code{176,22,{\leq}4}$ and
$\code{192,6,{\leq}8}$ entries.  The comparison set also includes five non-BB
weight-five entries: exact $\code{40,10,4}$, $\code{40,6,4}$, and
$\code{45,9,3}$ codes, and witness-only $\code{18,4,{\leq}3}$ and
$\code{676,71,{\leq}4}$ entries.

The basis-independent connectivity test covers all 102 entries in the
Challenge comparison set: 81 are connected, 12 decompose into multiple
pairwise-isomorphic positive-$k$ components, and nine decompose into one
positive-$k$ component together with $k=0$ stabilizer-state factors.  Submitted
logical witnesses were checked to lie in the retained component before distance
evidence was transferred.  The Challenge entries supply the short-block
points on the exact weight-six length--FOM curve: $\code{18,4,4}$ ($3.56$),
$\code{30,4,6}$ ($4.80$), $\code{48,4,8}$ ($5.33$), and
$\code{84,6,10}$ ($7.14$); all four are connected.  A weight-eight
$\code{154,8,{\leq}17}$ submission has verified logical
witnesses, setting the certified ceiling $\FOM\leq15.01$; the exact
$\code{144,14,14}$ code remains the preceding point on the length--FOM
frontier.  At weight four,
$\code{84,2,9}$ is connected, while the exact $\code{96,12,4}$ and
$\code{192,24,4}$ parents reduce to $\code{16,2,4}$ as described above; the
component has FOM $2.00$; the two remaining weight-four entries retain only
witness-based ceilings.  Among the five non-BB weight-five entries, the exact
$\code{40,10,4}$ code lies on the rate--distance frontier; the other two exact
entries and two witness-only entries do not change either exact Pareto frontier.
Within the exact subset of the compiled dataset, our constructions extend both
the weight-five distance envelope and its rate--distance frontier.  The
higher-weight cohorts retain the larger FOM envelope, defining a
stabilizer-weight tradeoff whose practical value depends on circuit and
hardware costs.

\subsection{PBB codes}
\label{sec:pbb-results}

The PBB search tested the support-containment ansatz in
Eq.~\eqref{eq:pbb}, with fitness downweighting estimated distances below five.
Exhaustive search certified every small-profile proposal.  Excluding the 106
with $d=2$, the retained 151
comprise 31 with $d=3$, 90 with $d=4$, and 30 with $d=6$.  Thus, 30 retained
proposals meet the $d\geq5$ target, while 121 have $d=3$ or 4.  These
counts are specific to the ansatz and search objective.  The largest
small-profile FOM is 1.50; one retained proposal attaining it has
parameters $\code{72,12,3}$.  Among the small-profile proposals meeting
the $d\geq5$ target, the $\code{144,4,6}$ parameter point has the highest FOM,
1.00, and is attained by five connected presentation classes.  Across both
profiles, reducing certified parents to connected components gives a largest
exact nonbaseline PBB FOM of 1.85, attained by two distinct connected
component classes with parameters $\code{216,4,10}$.

After excluding 25 exact $d=2$ proposals, the large profile contains 79
retained proposals with direct evidence for 54 exact distances and 25
certified intervals.  Combining evidence within presentation classes adds five
exact distances across three classes, for 59 exact proposals and 20
proposals in 14 interval classes.  Exhaustive search gives every interval
the lower endpoint $d\geq5$; their upper endpoints range from seven to twelve
across $n=288$ (2), $n=360$ (10), and $n=432$ (8).  Among the 25 initial
interval targets, 112 of 208 logical MILP instances exhausted their
per-instance time limits.
At the retained-proposal level, the largest exact distance is 11,
attained by a $\code{360,4,11}$ proposal, while the largest exact FOM is
1.85, attained by two $\code{432,8,10}$ proposals.  All three are
disconnected: the former decomposes into two copies of a $\code{180,2,11}$
component, and each latter proposal decomposes into two copies of its
respective $\code{216,4,10}$ component class.

We also give connected PBB presentations of these components.
The $\code{180,2,11}$ presentation matches a component of
$\code{360,4,11}$, while two $\code{216,4,10}$ presentations with distinct
colored-Tanner canonical labels match components of the two $\code{432,8,10}$
parent presentation classes.  A connected
$\code{144,4,8}$ presentation similarly matches a component of a certified
$\code{432,12,8}=3\times\code{144,4,8}$ parent.  Each distance transfer is
supported by an exact parent MILP certificate, a decomposition into
pairwise-isomorphic components, and matching canonical graph signatures for
the explicit connected presentation and a parent component.  Thus the largest
connected exact PBB FOM with $d\geq5$ is 1.85, attained by the two distinct
$\code{216,4,10}$ component classes, and the largest connected exact distance
is 11, attained by the displayed $\code{180,2,11}$ presentation.  Among
retained large-profile PBB proposals that are
themselves connected under the row-space analysis, the largest exact distance is
10, attained at both $n=360$ and $n=420$, and the largest exact FOM is 1.11,
attained by $\code{360,4,10}$.  The connected exact catalogue proposals
extend through $n=420$; the exact $n=432$ proposals reduce to smaller
connected components.  The proposals with parameters $\code{216,2,8}$,
$\code{360,4,10}$, and $\code{420,4,10}$ are members of a fixed baseline family
supplied independently of model output.

For reference, one of the two disconnected parent proposals with
parameters $\code{432,8,10}$ is defined on $(\ell,m)=(18,12)$ by
\begin{align}
A&=1+x^2y^2, & B&=1+xy^6+x^2,\nonumber\\
C&=1+x^2y^2, & D&=1+x^2.
\end{align}

All four explicit PBB component presentations fail the catalogue's tested
local-Clifford-to-CSS screens; classification under their complete
local-Clifford orbits remains open.  The shorter $\code{12,2,3}$ PBB point is
represented by four distinct component classes.  For these small codes, an
exhaustive search over all $6^{12}$ local Pauli-axis permutations proves that
none is local-Clifford-equivalent to CSS.  This point begins the compiled
weight-five length--FOM curve.

Table~\ref{tab:pbb} summarizes the selected PBB presentations and their evidence.

\begin{table*}[t]
\centering
\caption{Selected weight-five PBB presentations.  Exact rows are certified.  Interval
rows combine exclusion through weight four with a supported upper endpoint, so
their FOMs are upper bounds.  Evidence labels are Exh.+M (exhaustive search
corroborated by all-logical MILP), M (all-logical MILP), M+component (a
connected presentation matched by colored-BLISS to a component of a
homogeneous M-certified parent), and L4+I+B (certified $d\geq5$ plus both a
nonoptimal MILP incumbent and a decoder estimate).
Baseline rows identify fixed-family controls supplied by the campaign
evaluators independently of model output.  Each row specifies a particular
polynomial presentation, and repeated parameter triples need not denote the
same class.  In particular, the two $\code{216,4,10}$ rows belong to distinct
component classes, and the $\code{144,4,6}$ component class underlying the
displayed $\code{288,8,6}$ parent is distinct from that of the displayed
connected $\code{144,4,6}$ row.  ``conn.'' denotes connected.}
\label{tab:pbb}
\scriptsize
\setlength{\tabcolsep}{2pt}
\begin{tabular}{lcllcll}
\toprule
Parameters & $(\ell,m)$ & $(A,B)$ & $(C,D)$ & $\FOM$ & Structure & Evidence\\
\midrule
$\code{144,4,6}$ & $(9,8)$ & $(1+x^6y,\ 1+x^2y^4+x^4)$ & $(0,\ x^2y^4)$ & 1.00 & conn. & Exh.+M\\
$\code{144,4,8}$ & $(18,4)$ & $(1+x^3y,\ 1+xy^2+x^2)$ & $(1+x^3y,\ xy^2)$ & 1.78 & conn. & M+component\\
$\code{180,2,11}$ & $(15,6)$ & $(1+x^2y^2,\ 1+x^2y^3+x^4)$ & $(1+x^2y^2,\ 1+x^4)$ & 1.34 & conn. & M+component\\
$\code{216,2,8}$ & $(12,9)$ & $(1+x+x^2,\ 1+y)$ & $(x,\ 1+y)$ & 0.59 & conn. & M; baseline\\
$\code{216,4,10}$ & $(18,6)$ & $(1+x^2y,\ 1+xy^3+x^2)$ & $(1+x^2y,\ 1+x^2)$ & 1.85 & conn. & M+component\\
$\code{216,4,10}$ & $(18,6)$ & $(1+x^4y,\ 1+xy^3+x^2)$ & $(0,\ 1+x^2)$ & 1.85 & conn. & M+component\\
$\code{288,8,6}$ & $(12,12)$ & $(1+x^4y^2,\ 1+xy^6+x^2)$ & $(1+x^4y^2,\ 1+x^2)$ & 1.00 & $2\times\code{144,4,6}$ & M\\
$\code{288,4,5{\leq}d{\leq}8}$ & $(12,12)$ & $(1+x^3y,\ 1+xy^6+x^2)$ & $(1+x^3y,\ 1+x^2)$ & $\leq0.89$ & conn. & L4+I+B\\
$\code{360,4,11}$ & $(15,12)$ & $(1+x^2y^4,\ 1+x^2y^6+x^4)$ & $(1+x^2y^4,\ 1+x^4)$ & 1.34 & $2\times\code{180,2,11}$ & M\\
$\code{360,4,10}$ & $(15,12)$ & $(1+x+x^2,\ 1+y)$ & $(x,\ 1+y)$ & 1.11 & conn. & M; baseline\\
$\code{360,4,5{\leq}d{\leq}12}$ & $(15,12)$ & $(1+x^4y,\ 1+xy^6+x^2)$ & $(0,\ 1+x^2)$ & $\leq1.60$ & conn. & L4+I+B\\
$\code{420,4,10}$ & $(15,14)$ & $(1+x+x^2,\ 1+y)$ & $(x,\ 1+y)$ & 0.95 & conn. & M; baseline\\
$\code{432,12,8}$ & $(18,12)$ & $(1+x^3y^3,\ 1+xy^6+x^2)$ & $(1+x^3y^3,\ xy^6)$ & 1.78 & $3\times\code{144,4,8}$ & M\\
$\code{432,8,10}$ & $(18,12)$ & $(1+x^2y^2,\ 1+xy^6+x^2)$ & $(1+x^2y^2,\ 1+x^2)$ & 1.85 & $2\times\code{216,4,10}$ & M\\
$\code{432,8,10}$ & $(18,12)$ & $(1+x^4y^2,\ 1+xy^6+x^2)$ & $(0,\ 1+x^2)$ & 1.85 & $2\times\code{216,4,10}$ & M\\
$\code{432,4,5{\leq}d{\leq}12}$ & $(18,12)$ & $(1+x^2y^5,\ 1+xy^6+x^2)$ & $(0,\ 1+x^2)$ & $\leq1.33$ & conn. & L4+I+B\\
\bottomrule
\end{tabular}
\end{table*}

The PBB presentations selected in Table~\ref{tab:pbb} share a compact algebraic
form, which we call the
order-three geometric PBB subfamily.  Up to exchanging the two sectors, write
\begin{equation}
 \begin{aligned}
 A&=1+v, & B&=1+u+u^2,\\
 C&\in\{0,A\}, & D&\in\{u,1+u^2\}.
 \end{aligned}
\label{eq:pbb-geometric-family}
\end{equation}
Here $u$ and $v$ are lattice monomials.  Both
$AC^\mathsf{T}$ and $BD^\mathsf{T}$ are inversion-symmetric, so the PBB
commutativity condition holds identically.  Four parents in this ansatz yield
the explicit connected presentations above: the two distinct
$2\times\code{216,4,10}$ parents, the $3\times\code{144,4,8}$ parent, and the
$2\times\code{180,2,11}$ parent.  The same form also contains all four derived
presentations.  For selected $(A,B)$ pairs, the search evaluated all 31
nonempty support-contained $(C,D)$ choices; over the broader set, it sampled
only sparse perturbations.  Equation~\eqref{eq:pbb-geometric-family}
characterizes the retained subfamily; exhaustive enumeration of all
support-contained choices is a separate problem.  Each of the 33 connected
catalogue PBB presentation classes has an
underlying $(A,B)$ pair with an order-three quotient-character certificate.
Thus, all connected PBB classes sampled by the search share this order-three
backbone.  This common structure motivates an order-three geometric PBB
subfamily within the support-containment ansatz; broader sampling can test its
reach across non-CSS weight-five codes.

These four explicit PBB components connect this pattern to the CSS
classification above.  In each case, the supports of $A$ and $B$ lie along two
cyclic directions that jointly generate a noncyclic ambient group, while
$C$ and $D$ are supported on those same directions.  For $\code{144,4,8}$,
the two cyclic direction subgroups have orders 12 and 18 and intersect in a
subgroup isomorphic to $C_3$.  The corresponding orders are 15 and 30, with
intersection isomorphic to $C_5$, for $\code{180,2,11}$; for both
$\code{216,4,10}$ presentations, the orders are 18 and 18 and the intersection
is isomorphic to $C_3$.  The CSS
BB codes defined by the corresponding
$(A,B)$ pairs therefore belong to the same cyclic-support, cyclic-overlap
stratum, and the PBB terms are support-contained symplectic perturbations of
those codes.  A
lifted-product reduction would additionally require the relevant overlap
extensions to split, which is not implied by these support data alone.

The fixed baseline family is obtained from the sector-swapped form of
Eq.~\eqref{eq:pbb-geometric-family} by setting $u=x$ and $v=y$:
\begin{equation}
 (A,B,C,D)=(1+x+x^2,\ 1+y,\ x,\ 1+y).
 \label{eq:pbb-baseline-family}
\end{equation}
For $3\mid\ell$ and $m\geq2$, the dimension of this family is exactly
\begin{equation}
 k=2\gcd(m,2).
 \label{eq:pbb-baseline-dimension}
\end{equation}
To see this, put $N=\ell m$ and
$h=1+y+\cdots+y^{m-1}$.  The mixed top-row map has rank $N$ because
$C=x$ is a unit, whereas the CSS-type bottom-row map has rank $N-2$ because
$\operatorname{Ann}(A^*)\cap\operatorname{Ann}(B^*)$ has dimension two, where
$\operatorname{Ann}$ denotes the annihilator in $R_{\ell,m}$.
The two row spaces intersect in dimension two when $m$ is even and trivially
when $m$ is odd.  Indeed, the image of multiplication by $1+y$ is the kernel
of evaluation at $y=1$;
because $h(1)=m\bmod 2$, it contains $h$ precisely when $m$ is even.
Subtracting the stabilizer rank from $2N$ gives
Eq.~\eqref{eq:pbb-baseline-dimension}.

There is also a general analytic distance upper bound.  With
\[
 Q=(1+x)\sum_{j=0}^{\ell/3-1}x^{3j},
\]
the operators
\[
 L_x=(0,Q\mid0,Q),\qquad L_y=(0,0\mid0,h)
\]
are nontrivial logicals of Pauli weights $2\ell/3$ and $m$, respectively.
Here $AQ=0$ and $Bh=0$ establish commutation.  Nontriviality of $L_x$
follows because a stabilizer representation would require a coefficient
$p$ satisfying $pB=Q$, which is impossible after evaluating at $y=1$.
For $L_y$, a representation would require a coefficient $q$ satisfying
$qB=0$; because $D=B$, this also gives $qD=0$, leaving $rA^*=h$ for another
coefficient $r$.  Coefficientwise in $y$, this would make the zero divisor
$A^*$ invertible.  Consequently,
\begin{equation}
 d\leq\min\!\left(m,\frac{2\ell}{3}\right).
 \label{eq:pbb-baseline-bound}
\end{equation}

Every campaign lattice in this sequence has $m\geq2\ell/3$.  The exact
baseline-family presentations on these lattices saturate the $2\ell/3$ branch:
they include
$\code{72,4,4}$, $\code{120,4,4}$,
$\code{144,4,6}$,
$\code{216,2,8}$, $\code{288,4,8}$, $\code{360,4,10}$, and
$\code{420,4,10}$, together with the odd-$m$ members
$\code{108,2,4}$ and $\code{162,2,6}$.  The baseline $\code{144,4,6}$
presentation is distinct from the same-parameter connected row in
Table~\ref{tab:pbb}.  For the interval-certified baseline presentation on
$(\ell,m)=(18,12)$, which has parameters
$\code{432,4,5\leq d\leq12}$, Eq.~\eqref{eq:pbb-baseline-bound} supplies the
upper endpoint.  This presentation is likewise distinct from the
same-parameter row in Table~\ref{tab:pbb}.  Equality with the $2\ell/3$ branch
is not universal: the unsearched lattice $(\ell,m)=(18,2)$ has $n=72$, within
the stated block-length range, and gives the exact $\code{72,4,2}$ code that
saturates the $m$ branch.  This sequence is a
fixed baseline supplied independently of model output.  A released bit-level
verifier checks the stabilizer ranks, both logical constructions, and the
counterexamples~\cite{cruzbenito2026qcode}.

At $n=288$ and $n=432$, the interval ceilings 0.89 and 1.33 lie below the
largest exact FOMs at those same lengths, 1.00 and 1.85; the $n=360$ intervals
are the sole remaining possibilities for a same-length improvement.  Across
all lengths, every interval ceiling is at most 1.85, so the explicit
$\code{216,4,10}$ realizations define the connected PBB FOM envelope under the
current evidence.

Across the tested support-containment ansatz and parameter range, the certified
CSS frontier remains above every PBB candidate passing the implemented non-CSS
screens.  Broader non-CSS weight-five families provide a natural next test.
The two
weight-five mirror-code examples tabulated in Ref.~\cite{khesin2026mirror}
are equivalent to CSS codes under the authors' Hadamard test; the non-CSS
mirror-code constructions emphasized in that work have weight at least six.

\subsection{Weight-five transfer control}
\label{sec:uniform-results}

Table~\ref{tab:uniform-results} compares 6,444 logged positive-distance
campaign observations with three control replicates of 6,444 completed
selected evaluations.

\begin{table*}[tbp]
\centering
\caption{Weight-five transfer control.  Control entries are ranges over three
stratified-uniform replicates.}
\label{tab:uniform-results}
\small
\begin{tabular}{@{}p{0.32\textwidth}p{0.30\textwidth}p{0.30\textwidth}@{}}
\toprule
Endpoint & LLM-guided output & Stratified-uniform control\\
\midrule
Records/evaluations & 6,444 logged positive-distance observations &
  6,444 completed selected evaluations\\
Distinct proposals & 936 & 6,444\\
Distinct certified-$d\geq5$ proposals & 816/936 (87.18\%) &
  4,672--4,785/6,444 (72.50--74.26\%)\\
Connected certified-$d\geq5$ classes & 311 & 566--572\\
Highest-FOM campaign-certified class sampled & $\code{180,4,14}$; 4.356 &
  $\code{180,4,14}$; 4.356\\
\bottomrule
\end{tabular}
\end{table*}

At the distinct-proposal level, the campaign output has a 12.9--14.7
percentage-point higher certified-$d\geq5$ fraction.  The
without-replacement control replicates contain 6,444 distinct proposals versus
936 in the campaign output and therefore cover more connected certified
classes; this absolute coverage difference partly reflects the suppression of
repeats.

The control included low-weight certification through weight four but no
separate high-distance exact follow-up.  The last row therefore records
sampling of a presentation in a campaign-certified class, not the control's
maximum.  Of 385 pooled classes meeting the finalist rule (estimated
$\FOM\geq5$ and certified $d\geq5$), 359 remain without exact evidence; every
replicate samples the campaign-certified
$\code{180,4,14}$ class.  Each replicate evaluated the ranks of
94,775--96,487 sampled pairs to fill its quotas.  Because each stratum stops
when its quota is filled, $6{,}444$ divided by this count is not a full-space
sampling fraction.
Per-lattice and per-replicate summaries, all sampled draws and distance
records, the low-weight audits, and reproduction and validation scripts are
available in the code repository~\cite{cruzbenito2026qcode}.

\section{Limitations}
\label{sec:limitations}

The 630 presentation classes quotient symmetries of the stored generator
graphs, not arbitrary stabilizer-basis changes or general local-Clifford
operations, and are therefore an upper bound on code-equivalence classes.  The
PBB label ``non-CSS'' likewise refers to the tested transformation set, except
for the four $\code{12,2,3}$ component classes whose complete local-Clifford
orbits were checked.  The algebraic taxonomy in
Sec.~\ref{sec:algebraic-classification} also classifies the displayed
group-algebra presentations rather than all equivalent realizations.

The stratified-uniform experiment fixes the campaign's allocation by lattice
and dimension band and tests selection within those strata, not allocation
among them.  Its campaign denominator contains logged positive-distance
observations, whereas the control denominator contains all completed selected
evaluations.  Because failed and zero-distance calls are unavailable at the
required granularity, the reported campaign fraction can exceed its
attempted-call yield; the experiment therefore does not compare attempted-call
yield or total compute.  Campaign observations can also repeat
proposals, whereas the control samples are drawn without replacement, and the
control did not receive separate high-distance exact certification.
The broader uniform-random and genetic-algorithm benchmarks remain those of
our preceding work~\cite{cruzbenito2026evolutionary}.

The cyclic-kernel formula in Eq.~\eqref{eq:cyclic-kernel-dimension} and the
subgroup test make algebraic enumeration and symmetry reduction of the $2+3$
sector over cyclic ambient groups practical.  A complete census also requires
an exact distance or verified upper-bound witness for every surviving orbit;
this remains open over $C_{45}$, $C_{48}$, $C_{70}$, $C_{90}$, and $C_{210}$.

The release supports reproducing the analysis from the retained catalogue onward
through solver inputs, reconstructed logical witnesses, campaign observations,
and all uniform-control draws.  Repository documentation records operational
provenance and the limits of reconstructing model-generation history.

The present analysis covers algebraic code parameters, stabilizer connectivity,
and an ideal interaction-layer schedule.  Full syndrome-extraction circuits,
circuit-level noise, decoding thresholds, and logical-error rates form the next
layer of physical evaluation.

The connectivity analysis is complete for the retained weight-five catalogue
and every entry in the Challenge comparison set.  Earlier campaign entries
used in Fig.~\ref{fig:pareto-comparison} were reduced with the corresponding
translation-subgroup test wherever their polynomial supports were available; a
full row-space analysis of those earlier campaigns is outside the present
scope.

\section{Data and code availability}

The Supplemental Material gives a compact catalogue containing every proposal
in an exact class and one representative of each remaining nonexact class.  The
complete machine-readable catalogue and reproduction code are available in
the qcode-discovery repository~\cite{cruzbenito2026qcode}.  The repository also
contains campaign configurations, archived QEC Challenge comparison data,
certification inputs and witnesses, equivalence and connectivity analyses,
interaction schedules, uniform-control draws, and validation scripts.  A
manifest records configurations, versions, hashes, run metadata, and validation
history.

\section{Conclusion}

Building on our earlier LLM-guided program-evolution workflow, we discover
weight-five BB and PBB codes and subject the resulting catalogue to exact
distance certification and structural reduction.  The leading connected CSS
realizations are $\code{96,4,10}$, $\code{140,6,10}$, and
$\code{180,4,14}$, with FOMs 4.17, 4.29, and 4.356; the latter two lie on the
exact rate--distance frontier compiled here.  At parameter triples shared with
the Lin--Pryadko archive, we contribute exact certificates, explicit bivariate
presentations, and verified parent--component relations.  The comparison also
certifies seven archive constructions, including the $\code{132,4,12}$
construction over the nonabelian group $C_{11}\times S_3$, whose FOM of 4.364
is the highest compiled exact weight-five value.

The three-replicate weight-five transfer control reveals complementary search
behavior.  Of 936 distinct CSS proposals recorded with positive distance, 816
(87.18\%) are certified at $d\geq5$, compared with 4,672--4,785 of 6,444
proposals per stratified-uniform replicate (72.50--74.26\%).
Without-replacement sampling
covers more connected classes, and every replicate samples the
campaign-certified $\code{180,4,14}$ class.  This weight-five experiment
complements the broader uniform-random and genetic-algorithm benchmarks in our
earlier work.

The structural analysis identifies 409 of 1,142 retained proposals as
disconnected and reduces 630 presentation classes to 608 component classes.  A
parallel analysis of the 102-entry Unitary Foundation QEC Challenge comparison
set identifies $\code{16,2,4}$ as the connected component underlying two
weight-four frontier parents.  The algebraic analysis explains the absence of
$k=2$ in the sampled CSS family, organizes the connected classes into
common-kernel sectors associated with quotient characters of orders 3, 7, and
15, and yields a 28-class taxonomy of two-direction presentations within the
sampled lattice orders.

For PBB codes, the strongest connected exact parameter point is
$\code{216,4,10}$, attained by two distinct component classes, and a complete
group-level calculation proves that none of
the four $\code{12,2,3}$ component classes is local-Clifford-equivalent to CSS.
Across the investigated parameter range and current exact evidence, the CSS
discoveries define the stronger finite-length envelope.  The component results
motivate testing connectivity during generation and optimizing a canonical connected
component directly in future campaigns.  Together, the exact certificates,
structural reductions, and algebraic classification establish a reproducible
foundation for exhaustive enumeration over cyclic ambient groups, broader
equivalence testing, and circuit-level evaluation.

\section*{Acknowledgments}

The author thanks Andrew W.\ Cross and David Kremer for insightful
discussions during the conceptualization of this work and for their
comments throughout the drafting process.

OpenAI GPT-5.6 Sol (via Codex) and Anthropic Claude Opus 5 (via Claude Code)
assisted with manuscript drafting and data analysis.  The author takes
responsibility for all content and conclusions.

\bibliographystyle{apsrev4-2}
\bibliography{weight5_references}

\end{document}


\raggedbottom

\title{Supplemental Material for ``Large Language Model-Guided Discovery of
Weight-Five Bivariate Bicycle Codes''}

\author{Juan Cruz-Benito}
\affiliation{IBM Research}

\date{8 September 2026}

\maketitle

Throughout, BB denotes bivariate bicycle, PBB denotes perturbed bivariate
bicycle, and CSS denotes Calderbank--Shor--Steane.  We use BP--OSD for belief
propagation with ordered-statistics decoding, MILP for mixed-integer linear
programming, and FOM for the figure of merit $kd^2/n$.

\section*{Campaign and certification settings}

\begin{center}
\small
\begin{tabular}{lrrrrr}
\toprule
Campaign & Iterations & Population & Islands & Workers & Eval. timeout (s)\\
\midrule
CSS-small & 300 & 500 & 5 & 6 & 1,200\\
CSS-large & 300 & 750 & 6 & 24 & 900\\
PBB-small & 300 & 300 & 4 & 6 & 1,200\\
PBB-large & 300 & 450 & 5 & 16 & 1,800\\
\bottomrule
\end{tabular}
\end{center}

All campaigns used random seed 42, temperature 1.0, and model-selection
probabilities 0.49 for Claude Opus 5 and 0.51 for GPT-5.6 Sol.  CSS ranking
used 1,000 BP--OSD trials, three 500-trial refinements, 200 combination-sweep
ordered-statistics decoding (OSD-CS) trials, and a 300-s exact-search attempt;
post-search MILP allowed 1,800 s per logical objective and 14,400 s per code.
PBB ranking used low-weight enumeration, adaptive MILP budgets of 15/90,
30/180, or 60/360 s per logical objective/per code, and 1,000+500 BP--OSD
trials.  Initial post-search all-logical PBB MILP passes allowed 180 s per
logical objective for the small profile and 1,800 s for the large profile;
unresolved objectives were rerun with limits up to 3,600 and 14,400 s,
respectively.  Because the profiles used different time and parallelism
budgets, their output counts are descriptive rather than equal-budget yields.
Further configuration details are available in the code repository
~\cite{cruzbenito2026qcode}.


\setcounter{table}{0}
\renewcommand{\thetable}{S\arabic{table}}

\section{Weight-five catalogue and supporting audits}
\label{sec:w5-complete-catalogue}
\subsection{Catalogue construction}
The source logs contain 1,297 proposals (13,923 observations).  After duplicate observations are merged, the catalogue retains 1,142 proposals (12,036 observations), omitting exact $d\leq2$ rows and rows with a reported endpoint $d_{\rm rep}\leq2$.  Canonical labelling groups the retained normalized tuples into 630 presentation classes, defined here as stored-generator colored-Tanner isomorphism classes: 535 CSS and 95 PBB.  This classification captures sector exchanges, monomial shifts, and lattice automorphisms that preserve the colored generator graph.  It does not capture equivalences requiring stabilizer-basis changes or local Clifford operations, so 630 remains an upper bound on the number of inequivalent codes.  ID prefixes CS, CL, PS, and PL denote CSS-small, CSS-large, PBB-small, and PBB-large; the C or P label after the slash identifies the presentation class.  The four campaign tables print all 436 proposals in exact classes and one representative of each nonexact class (839 rows total); the complete 1,142-row catalogue and supporting evidence are available through the project repository~\cite{cruzbenito2026qcode}.

\subsection{Distance evidence and class merging}
Duplicate observations are merged by requiring exact values to agree, maximizing certified lower bounds, and retaining the smallest reported endpoint together with whether that endpoint is rigorous or decoder-only.  For one representative of each of 492 presentation classes containing all 858 proposals formerly lacking direct lower-bound evidence, we exhaustively searched both CSS logical sectors for nontrivial operators of weight at most four.  The search enumerates supports of at most two qubits and combines pairs whose syndromes cancel, rejecting combinations that are stabilizers.  The audit resolved 64 classes (94 proposals) at low weight: two $d=2$ proposals were omitted, and 62 classes (92 retained proposals) have $d=3$ or 4.  For the other 428 classes (764 proposals), it certified $d\geq5$.  Every retained row has either an exact distance or a certified positive lower bound paired with a reported endpoint.  Before the final class merge, 398 proposals have exact row-level status and 744 have unresolved row-level status.  Of the latter, 64 have rigorous upper bounds and 680 retain decoder estimates.  30 proposals with exact row-level status inherit an exact $d\leq4$ result from an audited class representative.  A separate set of 15 proposals with exact row-level status combines a transferred $d\geq5$ result with its own weight-five witness.  These transferred-evidence subsets measure a different property from the 45 proposals whose decoder estimates of 5 were matched by reconstructed and independently verified weight-five witnesses.

Merging distance evidence within each presentation class gives 227 exact classes and 403 classes with a certified lower bound and reported endpoint: 53 rigorous intervals and 350 lower-bound--plus--estimate records.  The exact classes cover 436 proposals: 398 already had exact row-level status before this merge and 38 acquire it by class isomorphism.  Class merging tightens without establishing an exact value for another 99 rows.  The campaign tables separate each proposal's own evidence from the class effect.  Where a class effect is shown, the displayed $d$ and FOM use the class-level evidence, whereas Evid. lists only evidence attached to that proposal.  The figure of merit is $\mathrm{FOM}=kd^2/n$; values are rounded to two decimals unless three are needed to distinguish a close comparison.

\subsection{Connectivity and component classes}
The basis-independent connectivity test finds that 409 of 1,142 retained proposals are disconnected, including 235 of the 398 proposals with exact row-level status.  By family, 307 of the 912 CSS proposals and 102 of the 230 PBB proposals are disconnected.  The component partition agrees with the generator-graph partition for every retained row.  At the class level, 268 of 630 classes are disconnected, including 166 of 227 exact classes.

The high-$k$ component audit resolves six retained direct sums: $10\times\code{42,6,3}$, $16\times\code{24,4,3}$, $12\times\code{36,4,3}$, $10\times\code{42,4,5}$, $8\times\code{48,4,3}$, and $8\times\code{48,4,6}$.  The seventh, $18\times\code{24,4,2}$, is excluded from the retained catalogue because $d=2$.

Selecting one connected component from each parent presentation and repeating the canonical labelling reduces 630 parent classes to 608 component classes: 528 CSS and 80 PBB, of which 206 are exact and 402 are nonexact.  Because this comparison uses presentation-graph isomorphism rather than complete stabilizer-code equivalence, 608 remains an upper bound on the number of inequivalent component codes.

\subsection{Local-Clifford screen}
A separate complete local-Clifford-to-CSS audit covers all four exact PBB component classes with parameters $\code{12,2,3}$.  An exhaustive branch-and-bound search over the $6^{12}$ assignments of Pauli-axis permutations to the 12 qubits shows that none is locally Clifford equivalent to CSS.  The machine-readable certificates are available through the project repository~\cite{cruzbenito2026qcode}.

\subsection{Table conventions}
Evidence codes are Exh. (exact exhaustive search), H$_c$ (exact component enumeration plus isomorphism and dimension additivity), M (MILP optimality for every logical sector), $L_w$ (exhaustive exclusion of logical operators through weight $w$), I (MILP incumbent), W (independently reconstructed and verified logical operator), and B (BP--OSD/decoder estimate; operator not retained).  For a nonexact row, $[L,U]$ denotes a certified lower bound and a rigorous upper bound, whereas $L;\widehat{U}$ pairs the certified lower bound with a decoder estimate; the FOM column uses the same notation.  Own evidence labels these cases Interval and Estimate, respectively.  Class effect identifies an exact value, a tighter endpoint, or a rigorous endpoint supplied by an isomorphic presentation.  Obs. is the number of raw-log observations, not an equivalence-class multiplicity.  Decomp. gives the connectivity decomposition computed from the displayed polynomials: \texttt{conn.} for one component, or $r\times\code{n_0,k_0}$ for $r$ pairwise-isomorphic disjoint components with parameters $\code{n_0,k_0}$.  Each component has the same true distance as the direct sum, so the row's exact distance or its certified lower bound and reported endpoint carry over unchanged.

\subsection{Generator attribution and artifacts}
Attr. uses G for GPT-5.6 Sol, O for Claude Opus 5, N for the seed/null source, F for the fixed PBB baseline evaluated in every run, and -- when no attribution record exists.  These labels record associated generator programs, not independent discovery or the distance-certification method.  At the proposal level, the retained CSS association counts are G=739 and O=313, with 181 in both sets, 35 seed/null-only rows, and six rows without an event.  For PBB-small, the attribution counts are 90 G, 51 G/O, 2 G/N, and 8 G/O/N; for PBB-large, they are 8 G, 56 G/O, 11 G/O/N, and 4 O.  F marks 20 retained fixed-baseline proposals and is orthogonal to the G/O/N partition: it can be appended to any model-association label and is not included in those partition counts.  Twenty-five PBB-large bursts remain ambiguous at the individual-program level, although their sets of associated models are fixed.

For the 51 retained PBB-large targets, the completed audit certifies 26 exact distances and leaves 25 ranges with certified lower bounds and supported upper endpoints.  Further details, the complete machine-readable catalogue, and supporting evidence are available through the project repository~\cite{cruzbenito2026qcode}.

\begingroup
\scriptsize
\setlength{\tabcolsep}{2pt}
\begin{longtable}{lrrrrrrrr}
\caption{Catalogue accounting after filtering.  ``Logged obs.'' counts all observations with a positive reported distance; ``Kept obs.'' restricts this count to retained proposals.  ``All'' and ``Kept'' give proposal counts before and after filtering, and ``Classes'' gives the number of retained presentation classes.  Exact and unresolved counts are proposal-level statuses before the final class merge and can include low-weight evidence transferred from a verified representative.  Unresolved rows pair a certified lower bound with either a rigorous upper bound or a decoder estimate.}\label{tab:w5-supp-summary}\\
\toprule
Campaign & Logged obs. & Kept obs. & All & Kept & Classes & Removed & Exact & Unresolved\\
\midrule
CSS-small & 5,610 & 5,269 & 382 & 365 & 164 & 17 & 99 & 266\\
CSS-large & 834 & 824 & 554 & 547 & 371 & 7 & 94 & 453\\
PBB-small & 4,890 & 3,625 & 257 & 151 & 39 & 106 & 151 & 0\\
PBB-large & 2,589 & 2,318 & 104 & 79 & 56 & 25 & 54 & 25\\
\midrule
Total & 13,923 & 12,036 & 1,297 & 1,142 & 630 & 155 & 398 & 744\\
\bottomrule
\end{longtable}
\endgroup

\begin{longtable}{llllll}
\caption{Explicit connected BB presentations of components of certified disconnected parents.  Canonical labelling verifies that the corresponding presentation graphs are isomorphic, transferring the component's certified distance to each presentation.}\label{tab:w5-explicit-components}\\
\toprule
Code & $(\ell,m)$ & $A$ & $B$ & FOM & Certified parent\\
\midrule
$\code{90,4,9}$ & $(5,9)$ & $1+x^{2}y^{8}+x^{3}y^{7}$ & $1+x^{2}y^{6}$ & 3.60 & $\code{180,8,9}$\\
$\code{96,4,10}$ & $(16,3)$ & $1+y+x^{2}y^{2}$ & $1+x^{3}$ & 4.17 & $\code{384,16,10}$\\
$\code{140,6,10}$ & $(5,14)$ & $1+y^{13}+x^{4}y^{11}$ & $1+xy^{7}$ & 4.29 & $\code{420,18,10}$\\
\bottomrule
\end{longtable}

\begin{longtable}{llllll}
\caption{Explicit connected PBB presentations of components of certified disconnected parents.  Canonical labelling verifies that the corresponding presentation graphs are isomorphic, transferring the component's certified distance to each presentation.}\label{tab:w5-explicit-pbb-components}\\
\toprule
Code & $(\ell,m)$ & $(A,B)$ & $(C,D)$ & FOM & Certified parent\\
\midrule
$\code{144,4,8}$ & $(18,4)$ & $(1+x^{3}y,\ 1+xy^{2}+x^{2})$ & $(1+x^{3}y,\ xy^{2})$ & 1.78 & $\code{432,12,8}$\\
$\code{180,2,11}$ & $(15,6)$ & $(1+x^{2}y^{2},\ 1+x^{2}y^{3}+x^{4})$ & $(1+x^{2}y^{2},\ 1+x^{4})$ & 1.34 & $\code{360,4,11}$\\
$\code{216,4,10}$ & $(18,6)$ & $(1+x^{2}y,\ 1+xy^{3}+x^{2})$ & $(1+x^{2}y,\ 1+x^{2})$ & 1.85 & $\code{432,8,10}$\\
$\code{216,4,10}$ & $(18,6)$ & $(1+x^{4}y,\ 1+xy^{3}+x^{2})$ & $(0,\ 1+x^{2})$ & 1.85 & $\code{432,8,10}$\\
\bottomrule
\end{longtable}

\begingroup
\renewcommand{\arraystretch}{0.92}
\setlength{\tabcolsep}{1pt}
\begin{longtable}{@{}TTTTTTTNNNN@{}}
\caption{Certified parameter summary after combining evidence within presentation classes.  For each exact parameter triple, Props. and Cls. respectively count retained proposals and presentation classes; Conn./disc. splits the classes into connected and disconnected.  The displayed distance and FOM reflect class-level evidence.  Rep. is chosen deterministically, preferring an exact proposal-level status and then a connected presentation.  The representative columns give its lattice, generators (with -- in $(C,D)$ for CSS), and component decomposition.  The selected representative need not supply the decisive certificate.  An asterisk marks a row containing both connected and disconnected classes; in such a row Decomp. still describes only the representative.  The seven explicit connected-component presentations in Tables~\ref{tab:w5-explicit-components} and \ref{tab:w5-explicit-pbb-components} are excluded from these counts.}\label{tab:w5-certified-summary}\\
\toprule
Fam. & Code & Rep. & $(\ell,m)$ & $(A,B)$ & $(C,D)$ & Decomp. & FOM & Props. & Cls. & Conn./disc.\\
\midrule
\endfirsthead
\multicolumn{11}{c}{\tablename\ \thetable\ (continued)}\\
\toprule
Fam. & Code & Rep. & $(\ell,m)$ & $(A,B)$ & $(C,D)$ & Decomp. & FOM & Props. & Cls. & Conn./disc.\\
\midrule
\endhead
\bottomrule
\endfoot
\bottomrule
\endlastfoot
CSS & $\code{120,4,4}$ & \texttt{CS-d74f23a3} & $(6,10)$ & $(1+y^{3},\,1+x+x^{2})$ & -- & conn. & 0.53 & 1 & 1 & 1/0\\
CSS & $\code{120,4,10}$ & \texttt{CS-07d0fcb6} & $(6,10)$ & $(1+xy+x^{5}y^{5},\,1+y)$ & -- & conn. & 3.33 & 6 & 1 & 1/0\\
CSS & $\code{120,8,3}$ & \texttt{CS-183fa7f4} & $(6,10)$ & $(1+xy^{2}+x^{2}y^{4},\,1+x^{3}y^{6})$ & -- & $2\times\code{60,4}$ & 0.60 & 2 & 1 & 0/1\\
CSS & $\code{120,8,4}$ & \texttt{CS-ab5fa0e9} & $(6,10)$ & $(1+x^{2}y^{2}+x^{4}y^{4},\,1+x^{3}y^{8})$ & -- & $2\times\code{60,4}$ & 1.07 & 5 & 3 & 0/3\\
CSS & $\code{120,8,5}$ & \texttt{CS-999229aa} & $(6,10)$ & $(1+y^{2},\,1+x+x^{5}y^{3})$ & -- & conn.\textsuperscript{*} & 1.67 & 17 & 9 & 3/6\\
CSS & $\code{120,16,3}$ & \texttt{CS-c600e4d1} & $(6,10)$ & $(1+x^{2}y^{2}+x^{4}y^{6},\,1+y^{2})$ & -- & $4\times\code{30,4}$ & 1.20 & 1 & 1 & 0/1\\
CSS & $\code{120,16,4}$ & \texttt{CS-9dfccec1} & $(6,10)$ & $(1+y^{4},\,1+x^{2}+x^{4}y^{4})$ & -- & $4\times\code{30,4}$ & 2.13 & 2 & 1 & 0/1\\
CSS & $\code{120,16,5}$ & \texttt{CS-435681c1} & $(6,10)$ & $(1+x^{2}y^{2}+x^{4},\,1+y^{6})$ & -- & $4\times\code{30,4}$ & 3.33 & 1 & 1 & 0/1\\
\end{longtable}
\endgroup

\begingroup
\setlength{\tabcolsep}{0.84pt}
\renewcommand{\arraystretch}{0.96}
\begin{longtable}{@{}TTTTTTTT@{\kern.5pt\vrule\kern.5pt}TTTTT@{}}
\caption{CSS-small compact catalogue (224 displayed of 365 retained proposals; 164 presentation classes).  The table includes every proposal in an exact class and one deterministic representative of every nonexact class.  Own evidence and Evid. describe the proposal before class merging; Class effect reports when merging makes the displayed $d$ and FOM exact, tighter, or rigorously supported.  Column abbreviations, filtering, and evidence conventions are defined above.}\label{tab:w5-supp-css-small}\\
\toprule
ID/class & $(\ell,m)$ & $\code{n,k}$ & $d$ & Own evidence & Class effect & FOM & $A$ & $B$ & Evid. & Obs. & Attr. & Decomp.\\
\midrule
\endfirsthead
\multicolumn{13}{c}{\tablename\ \thetable\ (continued)}\\
\toprule
ID/class & $(\ell,m)$ & $\code{n,k}$ & $d$ & Own evidence & Class effect & FOM & $A$ & $B$ & Evid. & Obs. & Attr. & Decomp.\\
\midrule
\endhead
\bottomrule
\endfoot
\bottomrule
\endlastfoot
\texttt{CS-435681c1/C040} & $(6,10)$ & $\code{120,16}$ & $5$ & Exact & -- & $3.33$ & $1+x^{2}y^{2}+x^{4}$ & $1+y^{6}$ & M & 9 & G & $4\times\code{30,4}$\\
\texttt{CS-9dfccec1/C038} & $(6,10)$ & $\code{120,16}$ & $4$ & Exact & -- & $2.13$ & $1+y^{4}$ & $1+x^{2}+x^{4}y^{4}$ & Exh. & 1 & G & $4\times\code{30,4}$\\
\texttt{CS-8146fb41/C038} & $(6,10)$ & $\code{120,16}$ & $4$ & Exact & -- & $2.13$ & $1+x^{2}y^{4}+x^{4}$ & $1+y^{6}$ & Exh. & 1 & G & $4\times\code{30,4}$\\
\texttt{CS-c600e4d1/C039} & $(6,10)$ & $\code{120,16}$ & $3$ & Exact & -- & $1.20$ & $1+x^{2}y^{2}+x^{4}y^{6}$ & $1+y^{2}$ & Exh. & 121 & G/O & $4\times\code{30,4}$\\
\texttt{CS-999229aa/C022} & $(6,10)$ & $\code{120,8}$ & $5$ & Exact & -- & $1.67$ & $1+y^{2}$ & $1+x+x^{5}y^{3}$ & L4+W+B & 2 & G & conn.\\
\texttt{CS-fc4dc8ca/C025} & $(6,10)$ & $\code{120,8}$ & $5$ & Exact & -- & $1.67$ & $1+y^{2}$ & $1+xy^{2}+x^{2}y^{6}$ & L4+W+B & 1 & G & $2\times\code{60,4}$\\
\texttt{CS-461d5f82/C031} & $(6,10)$ & $\code{120,8}$ & $5$ & Exact & -- & $1.67$ & $1+y^{4}$ & $1+xy+x^{5}y^{4}$ & L4+W+B & 7 & G & conn.\\
\texttt{CS-83ce21aa/C031} & $(6,10)$ & $\code{120,8}$ & $5$ & Exact & -- & $1.67$ & $1+y^{8}$ & $1+xy^{4}+x^{5}y^{7}$ & L4+W+B & 1 & O & conn.\\
\end{longtable}
\endgroup

\begingroup
\setlength{\tabcolsep}{0.84pt}
\renewcommand{\arraystretch}{0.96}
\begin{longtable}{@{}TTTTTTTT@{\kern.5pt\vrule\kern.5pt}TTTTT@{}}
\caption{CSS-large compact catalogue (391 displayed of 547 retained proposals; 371 presentation classes).  The table includes every proposal in an exact class and one deterministic representative of every nonexact class.  Own evidence and Evid. describe the proposal before class merging; Class effect reports when merging makes the displayed $d$ and FOM exact, tighter, or rigorously supported.  Column abbreviations, filtering, and evidence conventions are defined above.}\label{tab:w5-supp-css-large}\\
\toprule
ID/class & $(\ell,m)$ & $\code{n,k}$ & $d$ & Own evidence & Class effect & FOM & $A$ & $B$ & Evid. & Obs. & Attr. & Decomp.\\
\midrule
\endfirsthead
\multicolumn{13}{c}{\tablename\ \thetable\ (continued)}\\
\toprule
ID/class & $(\ell,m)$ & $\code{n,k}$ & $d$ & Own evidence & Class effect & FOM & $A$ & $B$ & Evid. & Obs. & Attr. & Decomp.\\
\midrule
\endhead
\bottomrule
\endfoot
\bottomrule
\endlastfoot
\texttt{CL-929c2a70/C248} & $(16,12)$ & $\code{384,64}$ & $3$ & Exact & -- & $1.50$ & $1+x^{4}y^{4}+x^{12}y^{8}$ & $1+x^{4}$ & H$_c$ & 1 & -- & $16\times\code{24,4}$\\
\texttt{CL-a25b2baf/C247} & $(16,12)$ & $\code{384,32}$ & $6$ & Exact & -- & $3.00$ & $1+x^{4}y^{2}+x^{4}y^{10}$ & $1+x^{2}y^{9}$ & H$_c$ & 1 & -- & $8\times\code{48,4}$\\
\texttt{CL-83dd7b5d/C245} & $(16,12)$ & $\code{384,32}$ & $5$ & Exact & -- & $2.08$ & $1+x^{2}y^{6}$ & $1+x^{2}y^{2}+x^{4}y^{4}$ & M & 2 & G & $8\times\code{48,4}$\\
\texttt{CL-a94de289/C244} & $(16,12)$ & $\code{384,32}$ & $4$ & Exact & -- & $1.33$ & $1+x^{2}y^{5}+x^{4}y^{10}$ & $1+x^{4}y^{6}$ & Exh.+I & 4 & G/O & $8\times\code{48,4}$\\
\texttt{CL-02bb1787/C243} & $(16,12)$ & $\code{384,32}$ & $4$ & Exact & -- & $1.33$ & $1+x^{4}$ & $1+y^{2}+y^{4}$ & Exh. & 2 & G/N & $8\times\code{48,4}$\\
\texttt{CL-d9f65a7e/C246} & $(16,12)$ & $\code{384,32}$ & $3$ & Exact & -- & $0.75$ & $1+x^{4}y^{2}+x^{10}y^{7}$ & $1+x^{2}y^{3}$ & H$_c$ & 1 & -- & $8\times\code{48,4}$\\
\texttt{CL-885dfa58/C233} & $(16,12)$ & $\code{384,16}$ & $10$ & Exact & -- & $4.17$ & $1+y^{4}+x^{2}y^{8}$ & $1+x^{3}y^{6}$ & M & 1 & O & $4\times\code{96,4}$\\
\texttt{CL-49103759/C232} & $(16,12)$ & $\code{384,16}$ & $8$ & Exact & -- & $2.67$ & $1+x$ & $1+x^{3}y^{4}+x^{7}y^{8}$ & M & 3 & G/O & $4\times\code{96,4}$\\
\end{longtable}
\endgroup

\begingroup
\setlength{\tabcolsep}{0.84pt}
\renewcommand{\arraystretch}{0.96}
\begin{longtable}{@{}TTTTTTTT@{\kern.5pt\vrule\kern.5pt}T@{\kern.5pt\vrule\kern.5pt}T@{\kern.5pt\vrule\kern.5pt}TTTTT@{}}
\caption{PBB-small compact catalogue (151 displayed of 151 retained proposals; 39 presentation classes).  The table includes every proposal in an exact class and one deterministic representative of every nonexact class.  Own evidence and Evid. describe the proposal before class merging; Class effect reports when merging makes the displayed $d$ and FOM exact, tighter, or rigorously supported.  Column abbreviations, filtering, and evidence conventions are defined above.}\label{tab:w5-supp-pbb-small}\\
\toprule
ID/class & $(\ell,m)$ & $\code{n,k}$ & $d$ & Own evidence & Class effect & FOM & $A$ & $B$ & $C$ & $D$ & Evid. & Obs. & Attr. & Decomp.\\
\midrule
\endfirsthead
\multicolumn{15}{c}{\tablename\ \thetable\ (continued)}\\
\toprule
ID/class & $(\ell,m)$ & $\code{n,k}$ & $d$ & Own evidence & Class effect & FOM & $A$ & $B$ & $C$ & $D$ & Evid. & Obs. & Attr. & Decomp.\\
\midrule
\endhead
\bottomrule
\endfoot
\bottomrule
\endlastfoot
\texttt{PS-5feb6ebb/P013} & $(6,6)$ & $\code{72,12}$ & $3$ & Exact & -- & $1.50$ & $1+x^{3}y^{3}$ & $1+xy^{3}+x^{2}$ & $0$ & $1+x^{2}$ & Exh. & 1 & G & $6\times\code{12,2}$\\
\texttt{PS-64d64fd9/P012} & $(6,6)$ & $\code{72,12}$ & $3$ & Exact & -- & $1.50$ & $1+x^{3}y^{3}$ & $1+xy^{3}+x^{2}$ & $0$ & $xy^{3}$ & Exh. & 8 & G/O & $6\times\code{12,2}$\\
\texttt{PS-106b8669/P014} & $(6,6)$ & $\code{72,12}$ & $3$ & Exact & -- & $1.50$ & $1+x^{3}y^{3}$ & $1+xy^{3}+x^{2}$ & $1+x^{3}y^{3}$ & $1+x^{2}$ & Exh. & 2 & G/N & $6\times\code{12,2}$\\
\texttt{PS-f4956d3c/P011} & $(6,6)$ & $\code{72,12}$ & $3$ & Exact & -- & $1.50$ & $1+x^{3}y^{3}$ & $1+xy^{3}+x^{2}$ & $1+x^{3}y^{3}$ & $xy^{3}$ & Exh. & 1 & G & $6\times\code{12,2}$\\
\texttt{PS-edb2da66/P013} & $(6,6)$ & $\code{72,12}$ & $3$ & Exact & -- & $1.50$ & $1+x^{3}y^{3}$ & $1+x^{4}+x^{5}y^{3}$ & $0$ & $1+x^{4}$ & Exh. & 1 & G & $6\times\code{12,2}$\\
\texttt{PS-1a20f6c1/P012} & $(6,6)$ & $\code{72,12}$ & $3$ & Exact & -- & $1.50$ & $1+x^{3}y^{3}$ & $1+x^{4}+x^{5}y^{3}$ & $0$ & $x^{5}y^{3}$ & Exh. & 3 & G & $6\times\code{12,2}$\\
\texttt{PS-686df635/P014} & $(6,6)$ & $\code{72,12}$ & $3$ & Exact & -- & $1.50$ & $1+x^{3}y^{3}$ & $1+x^{4}+x^{5}y^{3}$ & $1+x^{3}y^{3}$ & $1+x^{4}$ & Exh. & 1 & G & $6\times\code{12,2}$\\
\texttt{PS-66471642/P011} & $(6,6)$ & $\code{72,12}$ & $3$ & Exact & -- & $1.50$ & $1+x^{3}y^{3}$ & $1+x^{4}+x^{5}y^{3}$ & $1+x^{3}y^{3}$ & $x^{5}y^{3}$ & Exh. & 1 & G & $6\times\code{12,2}$\\
\end{longtable}
\endgroup

\begingroup
\setlength{\tabcolsep}{0.84pt}
\renewcommand{\arraystretch}{0.96}
\begin{longtable}{@{}TTTTTTTT@{\kern.5pt\vrule\kern.5pt}T@{\kern.5pt\vrule\kern.5pt}T@{\kern.5pt\vrule\kern.5pt}TTTTT@{}}
\caption{PBB-large compact catalogue (73 displayed of 79 retained proposals; 56 presentation classes).  The table includes every proposal in an exact class and one deterministic representative of every nonexact class.  Own evidence and Evid. describe the proposal before class merging; Class effect reports when merging makes the displayed $d$ and FOM exact, tighter, or rigorously supported.  Column abbreviations, filtering, and evidence conventions are defined above.}\label{tab:w5-supp-pbb-large}\\
\toprule
ID/class & $(\ell,m)$ & $\code{n,k}$ & $d$ & Own evidence & Class effect & FOM & $A$ & $B$ & $C$ & $D$ & Evid. & Obs. & Attr. & Decomp.\\
\midrule
\endfirsthead
\multicolumn{15}{c}{\tablename\ \thetable\ (continued)}\\
\toprule
ID/class & $(\ell,m)$ & $\code{n,k}$ & $d$ & Own evidence & Class effect & FOM & $A$ & $B$ & $C$ & $D$ & Evid. & Obs. & Attr. & Decomp.\\
\midrule
\endhead
\bottomrule
\endfoot
\bottomrule
\endlastfoot
\texttt{PL-ba29a9df/P040} & $(12,9)$ & $\code{216,2}$ & $8$ & Exact & -- & $0.59$ & $1+y$ & $1+x+x^{2}$ & $1+y$ & $x$ & M & 126 & G/O/N/F & conn.\\
\texttt{PL-39082ddc/P040} & $(12,9)$ & $\code{216,2}$ & $8$ & Exact & -- & $0.59$ & $1+x+x^{2}$ & $1+y$ & $x$ & $1+y$ & M & 126 & G/O/N/F & conn.\\
\texttt{PL-9b97a02b/P053} & $(12,12)$ & $\code{288,48}$ & $3$ & Exact & -- & $1.50$ & $1+x^{6}y^{6}$ & $1+x^{2}y^{6}+x^{4}$ & $1+x^{6}y^{6}$ & $1+x^{4}$ & Exh. & 20 & G/O & $24\times\code{12,2}$\\
\texttt{PL-78c3e7bf/P052} & $(12,12)$ & $\code{288,24}$ & $4$ & Exact & -- & $1.33$ & $1+x^{3}y^{6}$ & $1+x^{2}y^{6}+x^{4}$ & $0$ & $1+x^{4}$ & Exh. & 3 & G/O & $6\times\code{48,4}$\\
\texttt{PL-ad1cc828/P049} & $(12,12)$ & $\code{288,16}$ & $4$ & Exact & -- & $0.89$ & $1+xy$ & $1+x^{2}y^{6}+x^{4}$ & $0$ & $1+x^{4}$ & Exh. & 28 & G/O & $4\times\code{72,4}$\\
\texttt{PL-0bbae5c3/P048} & $(12,12)$ & $\code{288,16}$ & $4$ & Exact & -- & $0.89$ & $1+xy^{5}$ & $1+x^{2}y^{6}+x^{4}$ & $1+xy^{5}$ & $x^{2}y^{6}$ & Exh. & 13 & G/O & $4\times\code{72,4}$\\
\texttt{PL-d3deda3d/P048} & $(12,12)$ & $\code{288,16}$ & $4$ & Exact & -- & $0.89$ & $1+x^{3}y$ & $1+x^{2}y^{6}+x^{4}$ & $1+x^{3}y$ & $x^{2}y^{6}$ & Exh. & 15 & G/O & $4\times\code{72,4}$\\
\texttt{PL-aecf3364/P050} & $(12,12)$ & $\code{288,16}$ & $4$ & Exact & -- & $0.89$ & $1+x^{4}y^{2}$ & $1+x^{2}y^{6}+x^{4}$ & $1+x^{4}y^{2}$ & $x^{2}y^{6}$ & Exh. & 1 & G & $4\times\code{72,4}$\\
\end{longtable}
\endgroup

\bibliographystyle{apsrev4-2}
\bibliography{weight5_references}